\documentclass[12pt,letterpaper]{article}
\ifdefined\pdftexversion\pdfoutput=1\relax\fi
\usepackage{amsmath,amssymb,amsfonts,cite,xcolor,epsf,epsfig,epstopdf,graphics,booktabs,graphicx,mathtools,subcaption,physics,verbatim}
\def\thefootnote{*\arabic{footnote}}
\definecolor{ultramarine}{rgb}{0.07, 0.04, 0.56}
\definecolor{cadmiumgreen}{rgb}{0.0, 0.42, 0.24}
\definecolor{indigo(dye)}{rgb}{0.0, 0.25, 0.42}
\usepackage[linktocpage=true,breaklinks]{hyperref}
\hypersetup{
	colorlinks=true,
	citecolor=ultramarine,
	linkcolor=cadmiumgreen,
	urlcolor=indigo(dye),
}

\usepackage[utf8]{inputenc}

\numberwithin{equation}{section}

\usepackage{physics}
\usepackage{array}
\usepackage{tabularx}
\newcolumntype{P}[1]{>{\centering\arraybackslash}p{#1}}
\newcolumntype{Y}{>{\raggedright\arraybackslash}X}

\usepackage[shortlabels]{enumitem}

\usepackage{comment}

\usepackage{dcolumn}
\newcolumntype{M}[1]{>{\centering\arraybackslash}m{#1}}
\newcolumntype{N}{@{}m{0pt}@{}}

\usepackage[T1]{fontenc}
\usepackage{lmodern}
\usepackage{geometry}
\usepackage{amsthm}
\usepackage{bm}
\usepackage{microtype}
\usepackage{ulem}
\usepackage{tikz}
\usepackage{float}

\newcommand{\Mpl}{M_{\rm Pl}}

\newcommand{\D}{{\rm d}}

\newcommand{\be}{\begin{equation}}
	\newcommand{\ee}{\end{equation}}

\begin{document}
	
\begin{center}
		
\def\thefootnote{\fnsymbol{footnote}}
		
\vspace*{1.5cm}
{\Large {\bf Black holes and perturbations in F-UDHOST gravity}}
\\[1cm]
		
{Mohammad Ali Gorji$^{1}$, Simon Iteanu$^{2}$, Ekaterina Koreshkova$^{3}$, \\
Karim Noui$^{2}$, Pavel Petrov$^{1}$, Hugo Roussille$^{4}$}
\\[.7cm]
		
		{\small\textit{$^{1}$Cosmology, Gravity, and Astroparticle Physics Group, Center for Theoretical Physics of the Universe, Institute for Basic Science (IBS), Daejeon, 34126, Korea}}\\
		\vspace{0.25cm}
		{\small\textit{$^{2}$Universit\'e Paris-Saclay, CNRS/IN2P3, IJCLab, 91405 Orsay, France}}\\
		\vspace{0.25cm}
        {\small\textit{$^{3}$Institute for Nuclear Research of Russian Academy of Sciences, 117312 Moscow, Russia}}\\
		\vspace{0.25cm}
        {\small\textit{$^{4}$Institute for Theoretical Physics, Heidelberg University, Philosophenweg 16, 69120, Heidelberg, Germany}}
		
\end{center}
	
\begin{abstract}
We study static, spherically symmetric solutions in F-UDHOST gravity, a shift-symmetric nonlinear extension of UDHOST theories whose higher-derivative interactions are built from a scalar two-form. For a scalar field with a linear time dependence, the background equation governing the kinetic density reduces to the equation of motion of a simple one-dimensional point-particle system. Once this motion is determined, the metric follows by a single integration. Using this reduction, we construct several families of exact black hole solutions with scalar hair. The monopole perturbations reveal that for nonextremal black holes on the generic branch, at short wavelengths, the sound speed vanishes and the quadratic Killing Hamiltonian is unbounded from below. These results hint toward a strong coupling problem and instability of the system. A special family is not covered by this result. We study its monopole perturbations and determine conditions for positive Killing energy and real radial propagation speeds.
\end{abstract}
	
	\hrule
	\def\thefootnote{\arabic{footnote}}
	\setcounter{footnote}{0}
	
	\newpage

    \tableofcontents

\newpage

\section{Introduction}\label{sec:introduction}

General Relativity (GR) has been remarkably successful in its domain of validity, from solar system tests to gravitational wave observations~\cite{Will:2014kxa, Berti:2015itd, LIGOScientific:2026oim}. Nevertheless, several reasons suggest that GR may not give a complete description of gravity at all scales. Observationally, the dark matter problem and the late-time acceleration of the Universe point to missing ingredients in our description of gravity and cosmology. Theoretically, the cosmological constant problem \cite{Weinberg:1988cp,Peebles:2002gy,Padmanabhan:2002ji, Martin:2012bt} and the expectation that quantum gravitational effects become important at sufficiently high energies indicate that GR should be modified or embedded in a more complete framework. Among the many possible extensions of GR, scalar-tensor theories appear very naturally, as they provide one of the minimal ways of enlarging the gravitational sector by adding only one extra propagating degree of freedom.

Scalar-tensor theories have a long history, starting from Brans-Dicke theory \cite{Brans:1961sx}, and they arise in many different contexts (see~\cite{Kobayashi:2019hrl} and references therein). They include, for instance, DBI-type scalar models \cite{Silverstein:2003hf} and k-essence theories \cite{Armendariz-Picon:1999hyi,Garriga:1999vw,Armendariz-Picon:2000ulo}, which have been widely used in cosmology, in particular for inflation \cite{Chen:2006nt} and dark energy \cite{Copeland:2006wr}. In these theories, the action depends only on the scalar field and its first derivatives, so that the equations of motion contain at most second derivatives. One may then ask whether it is possible to extend this framework by including second derivatives of the scalar field already at the level of the action. This must be done with care, since a naive higher-derivative extension typically propagates an extra Ostrogradsky mode and hence leads to a ghost instability \cite{Woodard:2015zca}.

A major step in this direction was the construction of Horndeski theory \cite{Horndeski:1974wa}, which is the most general scalar-tensor theory built from the metric, the scalar field, and its second derivatives while keeping the equations of motion second order. For some time, the second-order nature of the equations of motion was viewed as the key requirement for avoiding Ostrogradsky ghosts. However, it was later realized that this requirement is too strong. In particular, after invertible disformal transformations, one can generate theories whose equations of motion are higher than second order while no extra ghost degree of freedom appears \cite{Zumalacarregui:2013pma,Gleyzes:2014dya,Gleyzes:2014qga}. This showed that second-order equations are sufficient, but not necessary, for avoiding an additional Ostrogradsky mode. This insight led to the construction of degenerate higher-order scalar-tensor (DHOST) theories \cite{Langlois:2015cwa,Langlois:2015skt,BenAchour:2016cay,BenAchour:2016fzp}, where the Lagrangian is organized order by order in powers of the second derivatives of the scalar field and suitable degeneracy conditions guarantee that the unwanted Ostrogradsky mode is removed. Disformal transformations also play a central role in the classification of DHOST theories \cite{BenAchour:2016cay,Crisostomi:2016tcp,Crisostomi:2016czh,Langlois:2018dxi}. Closely related ideas appear in mimetic gravity \cite{Chamseddine:2013kea}, which can be understood as arising from singular disformal transformations and has generated a large literature of its own \cite{Deruelle:2014zza,Gorji:2017cai,Langlois:2018jdg,Babichev:2021bim,Takahashi:2021ttd,Jirousek:2022rym,Jirousek:2022jhh,Babichev:2024eoh,Gorji:2025ajb,Gorji:2026lwk}.

More recently, it was pointed out that one may relax full degeneracy and still obtain a sensible theory if the would-be ghost mode can be safely integrated out. This viewpoint leads to the so-called unitary degenerate higher-order scalar-tensor (UDHOST) framework \cite{DeFelice:2018ewo,DeFelice:2021hps,DeFelice:2022xvq}, which can be regarded as a controlled extension of DHOST theories. In this context, the scordatura mechanism \cite{DeFelice:2022xvq,Motohashi:2019ymr,Gorji:2020bfl,Gorji:2021isn} plays a crucial role: a small and controlled breaking of exact degeneracy can regularize the theory and cure strong coupling problems, in close analogy with what happens in ghost condensation \cite{Arkani-Hamed:2003pdi}. It is also worth mentioning that scalar-tensor theories can be described in a very general way within spatially covariant formalisms \cite{Gao:2014soa,Gao:2014fra,Chen:2026hhl}, although full four-dimensional diffeomorphism invariance is not manifest there. From this perspective, UDHOST theories are different in spirit, since they start from a covariant four-dimensional Lagrangian (see also~\cite{Langlois:2020xbc}). Closely related to these developments is the effective field theory approach to cosmology \cite{Arkani-Hamed:2003pdi,Cheung:2007st,Creminelli:2008wc,Gubitosi:2012hu,Bloomfield:2012ff,Gleyzes:2013ooa,Gleyzes:2014rba,Frusciante:2019xia}, which provides a unifying language for scalar-tensor theories on cosmological backgrounds.

Most of the theories mentioned above are restricted to finite polynomial dependence on the second derivatives of the scalar field in the action. For instance, Horndeski theory contains terms up to cubic order in second derivatives \cite{Horndeski:1974wa}, while quadratic and cubic DHOST theories are organized up to second and third powers, respectively \cite{BenAchour:2016fzp}. More recently, scalar-tensor theories with higher powers, and even nonlinear functions, of scalar-derivative invariants have started to be explored. In \cite{Gorji:2026nes}, a covariant construction was developed using higher-derivative invariants built from gradients along hypersurfaces of constant scalar field. This construction provides a nonlinear extension of UDHOST theories and forms the basis of the present work. An overlapping class of extensions of UDHOST theories was constructed in~\cite{Michiwaki:2026xru} using a spatially covariant formulation.

Black hole and compact object solutions provide an important arena in which scalar-tensor theories can be tested. If a scalar-tensor theory is relevant on cosmological scales, for instance for inflation or dark energy, then at shorter distances it generically mediates an additional scalar force. Since such a force is strongly constrained in local tests of gravity, viable theories must either screen it, as in the Vainshtein mechanism (see~\cite{Kobayashi:2019hrl} and references therein), or make its effects sufficiently small. Compact objects, and black holes in particular, are therefore natural systems in which one may search for or constrain scalar-tensor effects. This is especially timely in view of the rapid progress in black hole observations, in particular through gravitational wave detections by LIGO/Virgo/KAGRA \cite{Creminelli:2017sry,Sakstein:2017xjx,Ezquiaga:2017ekz,Arai:2017hxj,Langlois:2017dyl,Kase:2018aps}. In many cosmological situations the scalar gradient is timelike, and for this reason black hole solutions with a scalar field linear in time have been studied extensively in recent years in Horndeski, DHOST, and related theories \cite{Babichev:2013cya,Kobayashi:2014eva,Ogawa:2015pea,Babichev:2016rlq,Tretyakova:2017lyg,Minamitsuji:2019shy,Tomikawa:2021pca,Minamitsuji:2022mlv,Mukohyama:2022enj,Khoury:2022zor,Bakopoulos:2023sdm,Langlois:2022ulw,Allahyari:2026rnm}.
The existence of these backgrounds does not, however, guarantee their perturbative stability. Black hole perturbations in Horndeski and DHOST theories have been studied using quadratic actions and linearized field equations~\cite{Kobayashi:2012kh,Kobayashi:2014wsa,Langlois:2021aji,Langlois:2022eta,Langlois:2022ulw}, as well as effective field theory methods~\cite{Mukohyama:2022enj,Khoury:2022zor,Mukohyama:2025owu,Takahashi:2026alb}. These studies have revealed ghost and gradient instabilities and strong coupling in several classes of hairy solutions~\cite{Ogawa:2015pea,deRham:2019gha,Takahashi:2021bml}, while the time evolution of monopole perturbations has also been investigated~\cite{Nakashi:2022wdg,Charmousis:2026bhe}.

In this paper, we study static, spherically symmetric solutions in a shift-symmetric nonlinear extension of UDHOST theories, which we call F-UDHOST. The scalar Lagrangian is $P(X,{\cal F})$, with arbitrary dependence on the scalar kinetic invariant $X$ and the higher-derivative invariant ${\cal F}$. The letter F refers to the scalar two-form ${\cal F}_{\mu\nu}$ defining ${\cal F}$, introduced in Sec.~\ref{sec-model}. 
For a scalar field with a linear time dependence $\phi=qt+\psi(r)$, the background equation governing the kinetic density reduces to the equation of motion of a simple one-dimensional point-particle system, with $X(r)$ playing the role of position and the radius playing the role of time. Once this motion is determined, the metric follows by a single integration. Using this reduction, we construct several families of exact black hole solutions with scalar hair, including Yukawa-like and power-law corrections to the Schwarzschild metric.
The monopole perturbations reveal a no-go theorem for nonextremal black holes on the regular generic branch.  At every finite radius in the connected static exterior, the short-wavelength monopole perturbation  has zero sound speed, and its quadratic Killing Hamiltonian is unbounded from below.  A special family with $P(X,{\cal F})=K(X)+Q(X){\vert{\cal F}\vert}^{1/2}$, where $K$ and $Q$ are functions of $X$, can evade this obstruction. For this family we derive the condition for nonnegative quadratic Killing energy and determine the two radial characteristics. 

The paper is organized as follows. In Sec.~\ref{sec-model} we define the theory, derive the point-particle reduction of its spherical field equations and discuss the analogy with nonlinear electrodynamics. We construct exact backgrounds in Sec.~\ref{sec:exact-solutions}. Section~\ref{sec:perturbations} presents the axial and monopole perturbations, establishes the generic no-go theorem and analyzes the special family that can evade it. We summarize the results in Sec.~\ref{sec:Summary}. Technical details of the covariant field equations and perturbation coefficients are given in the appendices.


\section{F-UDHOST and its symmetry-reduced dynamics}
\label{sec-model}
In this section, we introduce the F-UDHOST model and consider a static, spherically symmetric metric with a scalar profile linear in time, characterized by the parameter \(q\). We derive the equations of motion and describe the general strategy for constructing exact solutions. Finally, we discuss the analogy with nonlinear electrodynamics.

\subsection{The model: definition and some notations}

We consider the shift-symmetric F-UDHOST family, a subclass of the nonlinear higher-derivative scalar-tensor theories introduced recently in \cite{Gorji:2026nes}, defined by the action
\begin{align}\label{eq:action}
S[g_{\mu\nu},\phi] & = \int \D^4 x \, \sqrt{-g}\,
\left[
\frac{M_{\rm Pl}^2}{2}\, R
+ P\!\left(X,{\cal F}\right)
\right] \, .
\end{align}
We use the metric signature $(-,+,+,+)$. Here $R$ is the Ricci scalar, $M_{\rm Pl}$ the Planck mass and $P$ is, at this stage, an arbitrary function of the two invariants
\begin{align}\label{eq:X-B-def}
X \equiv g^{\alpha\beta}\phi_\alpha\phi_\beta ,
\qquad
{\cal F} \equiv XZ - Y^2 ,
\end{align}
with $Y$ and $Z$ being higher derivative invariants defined by
\begin{align}
Y \equiv \phi^\alpha X_\alpha ,
\qquad
Z \equiv X^\alpha X_\alpha.
\end{align}
Here, we are using the standard notations $\phi_\mu\equiv\nabla_\mu\phi$, $X_\mu\equiv\nabla_\mu X$, and similarly for higher derivatives of any tensors. Since $X_\mu$ contains second derivatives of $\phi$, the function $P(X,{\cal F})$ is a nonlinear higher-derivative scalar sector.

The combination ${\cal F}$ has a simple geometric interpretation when the scalar gradient is timelike. Indeed, when $X<0$, we can define the unit normal vector $n_\mu$ and the associated projector $\gamma_{\mu\nu}$ by
\begin{align}
	n_\mu=-{\phi_\mu}/{\sqrt{-X}},
	\qquad
	\gamma_{\mu\nu}=g_{\mu\nu}+n_\mu n_\nu \, .
\end{align}
Then, a direct calculation shows that 
\begin{align}
	{\cal F}=X\gamma^{\alpha\beta}X_\alpha X_\beta \, .
\end{align}
Thus ${\cal F}$ measures the foliation-spatial gradient of $X$, multiplied by $X$, which ensures the absence of Ostrogradski ghosts in the theory. Notice that although this interpretation of ${\cal F}$ assumes $X<0$, the algebraic definition ${\cal F}=XZ-Y^2$ itself remains meaningful independently of this foliation interpretation. 

{In the quadratic DHOST basis of \cite{BenAchour:2016fzp}, the invariant ${\cal F}$ takes the form
\begin{equation}
    {\cal F}=4\Big(XL_4^{(2)}-L_5^{(2)}\Big)\,, \qquad L_4^{(2)}\equiv\phi^\mu\phi_{\mu\rho}\phi^{\rho\nu}\phi_\nu \, , \quad L_5^{(2)}\equiv(\phi^\mu\phi_{\mu\nu}\phi^\nu)^2
\, , \end{equation}
since $Z=4L_4^{(2)}$ and $Y^2=4L_5^{(2)}$. For the subclass linear in ${\cal F}$, the coefficients in the conventions of~\cite{DeFelice:2022xvq} are
\begin{equation}
    A_1=A_2=A_3=0\,,\qquad
    A_4=4XP_{\cal F}\,,\qquad A_5=-4P_{\cal F}\,,
\end{equation}
with Ricci curvature coefficient $\Mpl^2/2$ and no $\Box\phi$ interaction (known as the cubic Galileon). Consequently, following the notations of  \cite{DeFelice:2022xvq}, only one degeneracy condition is satisfied since
\begin{equation}
    c_U=c_{\mathrm D1}=0\,,\qquad
    c_{\mathrm D2}=4X^3P_{\cal F}\,.
\end{equation}
Therefore, when $P_{\cal F}\neq0$ the full action~\eqref{eq:action} is UDHOST rather than DHOST, since only the first degeneracy condition  is satisfied. Full DHOST degeneracy would additionally require $c_{\mathrm D2}=0$. This subclass lacks the scordatura $k^4$ contribution controlled by $c_{\mathrm D1}$ in the dispersion relation around stealth backgrounds.}

Before going further, it is also interesting to recall \cite{Gorji:2026nes} that ${\cal F}$ can equivalently be reformulated in terms of a two-form ${\cal F}_{\mu\nu}$ as follows
\begin{eqnarray}
\label{calFforscalar}
{\cal F}=\frac{1}{2}{\cal F}{}\!_{\alpha\beta} {\cal F}^{\alpha\beta} \, , \qquad {\cal F}{}\!_{\mu\nu} = \phi_\mu X_\nu - \phi_\nu X_\mu \, ,
\end{eqnarray}
which motivates the name F-UDHOST for the full $P(X,{\cal F})$ family, including arbitrary explicit dependence on $X$. Its $P({\cal F})$ subclass admits the analogy with nonlinear electrodynamics discussed in Sec.~\ref{NLEDanalogy}.

\subsection{Symmetry reduced configurations}

 We now consider  static and spherically symmetric solutions which correspond to solutions of the form
\begin{align}\label{eq:spherical background}
	{\D}s^2 &= - f(r)\,{\D}t^2
	+ \frac{{\D}r^2}{h(r)}
	+ r^2 \D\Omega^2,
	\qquad
	\phi = q\, t + \psi(r),
\end{align}
where \(\D\Omega^2 ={\D}\theta^2 + \sin^2\theta\, {\D}\varphi^2\) corresponds to the canonical metric on the unit 2-sphere. We have introduced $f$, $h$ and $\psi$ as functions of $r$ only while $q$ is a constant. As it is well-known, the shift symmetry of the action, i.e. its invariance under the transformation \(\phi\to\phi+c\) with \(c\) an arbitrary constant, makes the fact that the scalar field depends linearly on time is compatible with a static, spherically symmetric geometry~\cite{Babichev:2013cya}.

For the background configuration \eqref{eq:spherical background}, the two invariants \(X\) and \({\cal F}\) reduce to
\begin{align}\label{eq:X-B-SS}
	X= - \frac{q^2}{f} + h\, \psi^{\prime\;2},
	\qquad 
	{\cal F}= - q^2\frac{ h}{f} X'^2 \, ,
\end{align}
where the prime denotes the derivative with respect to $r$. Hence,  as we are going to confirm in the following section, the scalar sector can be  entirely described in terms of \(X\)  in this ansatz in the sense that the equations of motion can be formulated in term of $X$ and its derivatives.

\subsection{Equations of motion}

The covariant  metric field equations following from the action \eqref{eq:action} are given by 
	\begin{align}
    \label{Einsteinequationstensortext}
		{\cal E}_{\mu\nu} \equiv \Mpl^2 G_{\mu \nu } 
		-   T_{\mu\nu} \; = \; 0 \, , 
    \end{align}
    with the effective stress-energy tensor defined by
    \begin{align}
          T_{\mu\nu} = P g_{\mu\nu} - 2 \left[
    \mathcal Q \,  \phi_\mu \phi_\nu -2 Y P_{\cal F} \, \phi_{(\mu} X_{\nu)} + X P_{\cal F} \, X_\mu X_\nu 
    \right] \, .
        \label{effectivestress}
	\end{align}
The definition of $\mathcal Q$ and some details of the calculation can be found in Appendix~\ref{app:field-equations}.  
We are using the notation $P_{\cal F}$ for the derivative of $P$ with respect to ${\cal F}$. 
We now
specialize these equations to static and spherically symmetric configurations \eqref{eq:spherical background} in which case only three components of the Einstein equations \eqref{Einsteinequationstensortext} are non-vanishing. As usual, the scalar equation
is not independent of the Einstein equations due to the invariance under diffeomorphisms and it can be recovered taking the divergence of \eqref{Einsteinequationstensortext} and using Bianchi identity, i.e.
\begin{eqnarray}
    {\cal E}_\phi \; \propto  \; \phi_\beta \nabla_\alpha T^{\alpha\beta} \, .
\end{eqnarray}
As it is well-known, this equation is equivalent to the conservation of the  Noether current associated to the invariance of the action under shift-symmetry (see Appendix~\ref{app:field-equations}). 

First of all, we consider the non-diagonal non-trivial component of the Einstein equations ${\cal E}_{tr}=0$. Since the corresponding Einstein tensor $G_{tr}$ vanishes identically for static and spherically symmetric metric \eqref{eq:spherical background}, this equation reduces to $T_{tr}=0$ which leads to 
\begin{eqnarray}
\label{eqforQ}
   q \psi'(\mathcal Q - h P_{\cal F} X'^2) = 0 \, .
\end{eqnarray}
 In the generic branch where \(q\psi'\neq0\), this equation fixes $\mathcal Q=h P_{\cal F} X'^2$ which in turn implies, after a direct calculation,
\begin{align}
\label{TttTrr}
    T^t{}_{t} \; = \; T^r{}_{r} \, .
\end{align}
The remaining equations are the two diagonal non-trivial components of the Einstein equations. Using \eqref{TttTrr} on one side and the purely geometric identity (for static and spherically symmetric space-times),
\begin{align}
    G^t{}_{t}-G^r{}_{r}
    =\frac{h}{r}\left(\frac{h'}{h}-\frac{f'}{f}\right) \, ,
\end{align}
we immediately see that, as in GR, 
\begin{align}
\label{eq:f-equals-h}
	{\cal E}_{t}{}^t - {\cal E}_{r}{}^r = 0 \; \Longrightarrow \quad   f/h \; = \; \text{cte} \, .
\end{align}
 Without loss of generality, we can fix the integration constant to $1$ by a constant rescaling of the time coordinate. We use the same symbol $q$ for the scalar slope in the rescaled coordinate. Hence, the metric reduces to a single function, and $f=h$ in the following background solutions. 
 
Finally, the time component of the Einstein equations ${\cal E}_{tt}=0$ gives the first-order  equation for the function $f$, 
\begin{align}\label{eq:metric-master}
	\frac{\Mpl^2}{r^2} 
	\frac{\D}{\D{r}}\left[r(f-1)\right] = P-2{\cal F} P_{{\cal F}} \, .
\end{align}
In conclusion, the system is completely determined by the two equations \eqref{eqforQ} and \eqref{eq:metric-master}, together with the condition $f=h$ given in \eqref{eq:f-equals-h}. Solving the system therefore reduces to determining the functions $f(r)$ and $X(r)$, up to integration constants.

Interestingly, using $f=h$, one can show that $\mathcal Q$ reduces to 
\begin{eqnarray}
  \mathcal Q= P_X + Z P_{\cal F} + \nabla_\alpha H^\alpha   = P_X+ fX'^2P_{\cal F} + \frac{2q^2}{r^2} \frac{\D}{\D{r}}\left(
	r^2  X' P_{\cal F}
	\right) \, ,
\end{eqnarray}
so that, using  \({\cal F}=-q^2X'^2\), one can recast  \eqref{eqforQ} as the  Euler-Lagrange equation
\begin{align}\label{eq:SF-master}
	\frac{1}{r^2}\frac{\D}{\D{r}}\left(
	r^2 \,\frac{\partial P}{\partial X'}
	\right)
	=
	\frac{\partial P}{\partial X} \, .
\end{align}
Whenever derivatives with respect to $X'$ occur, $P$ denotes the radial function $P(X,-q^2X'^2)$, with $X$ and $X'$ treated as independent variables, i.e. 
\begin{eqnarray}
    \frac{\partial P}{\partial X'} = -2q^2 X' \, \frac{\partial P}{\partial \cal F} \, .
\end{eqnarray}
Derivatives with respect to ${\cal F}$ always refer to the covariant function $P(X,{\cal F})$. The scalar equation is therefore the Euler-Lagrange equation of a point particle in one dimension, with coordinate $X$, evolution parameter $r$, and Lagrangian $r^2P(X,-q^2X'^2)$.

In fact,~\eqref{eq:SF-master}, together with the metric equations \eqref{eq:f-equals-h} and \eqref{eq:metric-master}, can be derived directly from the reduced action obtained by imposing the static and spherically symmetric ansatz \eqref{eq:spherical background} in the original covariant action \eqref{eq:action}, provided that $X$ is treated as an independent field, namely,
\begin{eqnarray}
    S_{\rm red}[f,h,X] = \int \D{r} \sqrt{\frac{f}{h}}\left[ \Mpl^2 
    (1-h-r h') + r^2 P(X,{\cal F})\right] \, ,
\end{eqnarray}
where ${\cal F}$ is given by \eqref{eq:X-B-SS}. Constant time and angular factors, as well as boundary terms, are omitted in this radial action. An easy calculation of the associated equations of motion leads to the same three equations \eqref{eq:f-equals-h}, \eqref{eq:metric-master} and \eqref{eq:SF-master}.

\subsection{General strategy for the resolution}

The  strategy to find solutions is now simple. After choosing \(P\), one solves \eqref{eq:SF-master} for \(X(r)\), substitutes the result into \eqref{eq:metric-master}, and integrates for \(f(r)\) as follows
\begin{eqnarray}
    f(r)=1- \frac{2 }{\Mpl^2}  \frac{m(r)}{r}\, , 
\end{eqnarray}
where the ``mass'' function $m(r)$ is defined up to a constant by the integral
\begin{eqnarray}
 m(r) \; = \;  \frac{1}{2}  \int^r  \D x \; x^2 \left[  X'(x) {P}_{X'}(x) - P(x) \right] \,
 \label{massfunction}
\end{eqnarray}
where $P(x)$ holds for $P(X(x),{\cal F}(x))$. 
Once \(X(r)\) and \(f(r)\) are known, the radial scalar profile follows from
\begin{align}\label{eq:psi-reconstruction}
	f^2\psi'^{\,2}= f X+{q^2}  \, ,
\end{align}
which implies that the solution for $\psi$ is real  and then physically relevant when 
\begin{align}\label{eq:psi-reality}
	{f X+q^2}\ge0 .
\end{align}
On an asymptotically flat exterior branch with \(f\to1\) when $r \to \infty$, a vanishing asymptotic radial gradient therefore requires \(X_\infty=-q^2\). 

\subsection{Analogy with nonlinear electrodynamics}
\label{NLEDanalogy}

Interestingly, the above construction exhibits a strong
analogy with the well-known framework of Einstein gravity  coupled to nonlinear
electrodynamics (NLED) \cite{BornInfeld1934,Sorokin:2021tge,
AyonBeatoGarcia1998, Bronnikov2001}, whose dynamics is governed by the action
\begin{equation}
S[g_{\mu\nu},A_\mu]=\int \D^4x\,\sqrt{-g}
\left[
\frac{M_{\rm Pl}^2}{2}R
+\mathcal{L}(F)
\right],
\end{equation}
where $\cal L$ is a generic function of the electromagnetic invariant
\begin{equation}
F\equiv -\frac14 F_{\alpha\beta}F^{\alpha\beta},
\end{equation}
with $F_{\mu\nu}= \partial_\mu A_\nu-\partial_\nu A_\mu$ the field strength associated with the $U(1)$ gauge field $A_\mu$. 
This action presents analogies with the F-UDHOST model~\eqref{eq:action} in the particular case where $P$ depends on ${\cal F}$ only. In that case, the two-form ${\cal F}_{\mu\nu}$ \eqref{calFforscalar} plays a role analogous to that of $F_{\mu\nu}$ with the important difference that ${\cal F}_{\mu\nu}$ is not the curvature of a gauge field anymore. This difference leads to  different equations of motion for the gauge field in NLED and for the scalar field in F-UDHOST gravity. In the former case, the gauge field satisfies a generalized Maxwell equation
\begin{align}
\nabla_\alpha\left(\mathcal{L}_F F^{\alpha\mu}\right)=0,
\end{align}
where $\mathcal{L}_F$ denotes the derivative of $\cal L$ with respect to $F$, while in the latter  case, we have the more involved scalar equation
\begin{align}
\nabla_\alpha J^\alpha=0 \, , \qquad J^\mu = - P_{\cal F} X_\alpha {\cal F}^{\alpha\mu} - 2 \phi^\mu \nabla_\beta (P_{\cal F} \phi_\alpha {\cal F}^{ \alpha\beta}) \, ,
\end{align}
which involves the projections of the two-form ${\cal F}_{\mu\nu}$ in the directions $\phi^\mu$ and $X^\mu$. 

However, when one restricts the solutions to static and spherically symmetric configurations, the analogy between the two theories becomes stronger. If in addition, we suppose that the gauge field is purely electric, i.e. its only non-vanishing component is $A_0(r)$, the generalized Maxwell equation simplifies drastically and the system admits a first integral $Q_e$, the conserved electric charge, which enters in the solution as follows:
\begin{align}
r^2\mathcal{L}_F E=Q_e ,\qquad E\equiv\sqrt{\frac hf}\,A_0' .
\end{align}
Here $E$ is the norm of the electric field, i.e. $E_\mu E^\mu = E^2$. 
This equation completely
determines the electric field once the nonlinear electromagnetic Lagrangian
$\mathcal{L}(F)$ is specified since $F=E^2/2$ in the static and spherically symmetric configuration. 

Likewise, in the $P({\cal F})$ subclass of F-UDHOST gravity,
the equation for the scalar field simplifies in a very similar way. Indeed, first the only non-trivial components of the two-form are ${\cal F}_{0r}=-{\cal F}_{r0}= q X'$, second the radial component of the current $J^r$ vanishes due to the non-diagonal component of the Einstein equations, and as a consequence, using $f=h$, the scalar equation reduces to \begin{eqnarray}
    J^r = \frac{2 q^2 f \psi'}{r^2}\,\frac{\D}{\D r}\left[r^2 P_{\cal F}X'\right] \; = \; 0 \, .
\end{eqnarray} 
As expected, we recover \eqref{eq:SF-master} or equivalently \eqref{eqforQ}, which admits a first integral $Q_s$, namely
\begin{eqnarray}
\label{Qsintegral}
r^2  P_{\cal F} X' =Q_s,
\end{eqnarray}
and, following the analogy with NLED, we can interpret $Q_s$ as a scalar charge. Hence,  such an equation enables us to
completely determine $X'$ and then deduce the radial scalar profile of the kinetic density $X(r)$. 

As announced above, there is a clear analogy between the $P({\cal F})$ subclass of F-UDHOST gravity and NLED coupled to Einstein gravity: the scalar gradient $X'$ plays the role of the electric field $E$ (more precisely $E \leftrightarrow -qX'$), while ${\cal F}$ is the analogous of $F$, up to irrelevant constants.

The analogy extends to the gravitational sector since, in NLED coupled to gravity, the identity $T^t{}_t=T^r{}_r$ still holds, hence we recover the identity $f=h$ for static and spherically symmetric
solutions with  the metric equation
\begin{eqnarray}
\frac{M_{\rm Pl}^2}{r^2} \frac{\D}{\D r} \!\left[r(f-1)\right]
=
\mathcal{L}-2F\mathcal{L}_F \, ,
\end{eqnarray}
which is strictly equivalent to \eqref{eq:metric-master}. Indeed, 
the quantity
$P-2{\cal F}P_{\cal F}$ plays exactly the same role as the combination
$\mathcal{L}-2F\mathcal{L}_F$ that governs static black hole solutions in
Einstein-NLED theories.

Therefore, just as in nonlinear electrodynamics, the complete system reduces
to two coupled first-order ordinary differential equations: one determining
the matter sector (the electric field in NLED or the scalar profile in the
present theory), and the other determining the metric. This structural
equivalence suggests that the present scalar theory can naturally be viewed as
a scalar counterpart of Einstein-NLED models.

\section{Exact solutions: examples and properties}
\label{sec:exact-solutions}
This section is devoted to the the study of solutions to the equations of motion. We start with the class of theories that can be viewed as scalar analogues of nonlinear 
electrodynamics, characterized by a function $P$ depending solely on ${\cal F}$. Then,
we  consider the more general case with $P(X,{\cal F})$, and provide 
explicit examples of theories which admit exact solutions with interesting properties.

\subsection{Scalar nonlinear electrodynamics analogue: the \texorpdfstring{$P({\cal F})$}{P(F)} subclass}
\label{sec:Theories-with-B-only}
In this subsection we restrict the F-UDHOST family to $P=P({\cal F})$. This subclass provides the scalar analogue of nonlinear electrodynamics discussed in Sec.~\ref{NLEDanalogy}.
In that case, the equation for $X$ simplifies drastically since there exists a first integral $Q_s$ and \eqref{Qsintegral} can be reformulated as
\begin{eqnarray}
    P_{X'}(X'(r)) = \frac{\lambda}{r^2} \, , \qquad \lambda=- {2q^2 Q_s} \, .
\end{eqnarray}
We can therefore, in principle, solve $X'$ as a function of $r^2/\lambda$. More precisely, if $P_{X'}$ has a nonzero derivative in an open set the previous equation can be locally inverted by virtue of the implicit function theorem  and $X'$ is (locally) given by
\begin{eqnarray}
    X'(r)= g (\lambda/r^2) \, ,
\end{eqnarray}
where $g$ is the local inverse function of $P_{X'}$, i.e. $P_{X'} \circ g (u) = u$. Showing the existence of a global solution is more involved, but is not needed here. Still locally, we easily find that
\begin{eqnarray}
    P-X'P_{X'}   =  
    C_0 - \Pi(\lambda/r^2) \,,
\end{eqnarray}
where we have introduced the notations 
\begin{eqnarray}
  C_0= P(r_0) - X'(r_0) P_{X'}(r_0) \, , \qquad  \Pi(x) = \int_{x_0}^x g(u) \, \D u \, , \qquad 
  x_0=\lambda/r_0^2 \, .
\end{eqnarray}
This leads to the following integral expression for the metric coefficient:
\begin{eqnarray}
    f(r) = 1-\frac{2m_0}{\Mpl^2r}
    +\frac{1}{\Mpl^2r}\int^r \D x\,x^2\left[C_0-\Pi(\lambda/x^2)\right] \, ,
\end{eqnarray}
where $m_0$ is an integration constant. Additive constants in the radial primitives are absorbed into $m_0$. If the function $\Pi$ is defined for arbitrarily large values of $r$ and admits a finite limit $\Pi(0)$ as $r\to\infty$, the same solution can be written as
\begin{equation}
\begin{aligned}
 f(r)&=1-\frac{2m_0}{\Mpl^2r}-\frac{\Lambda r^2}{3}
 -\frac{1}{\Mpl^2r}\int^r\D x\,x^2
 \left[\Pi(\lambda/x^2)-\Pi(0)\right]\,,
\end{aligned}
\end{equation}
with an effective cosmological constant defined by
\begin{equation}
\begin{aligned}
 \Lambda&\equiv\frac{\Pi(0)-C_0}{\Mpl^2}\,.
\end{aligned}
\end{equation}
As a consequence, given a theory with $P$, one can easily obtain the solution for $f$. The inverse problem is also interesting. Indeed, one can ask whether it is possible to find a function $g$ which enables us to produce a given  solution $f$. This answer is positive if there exists $g$ such that
\begin{eqnarray}
   g(\lambda/r^2) =  \frac{\Mpl^2r^3}{2\lambda} \frac{\D}{\D r}\left[\frac{1}{r^2} \frac{\D}{\D r}[r(f-1)]\right] \, . 
\end{eqnarray}

To make the solutions more explicit, we consider a family of examples in which one can exactly compute the functions $g$.    In the case where $P$ is a power (with a constant) of the form
\begin{eqnarray}
    P(X')=-\Mpl^2\Lambda + \frac{b}{n+1} X'^{n+1} \, ,
\end{eqnarray}
where $n \neq -1$ is not necessarily an integer, then $P_{X'}=b X'^n$ and the solution for $X'$ reads
\begin{eqnarray}
    X'(r)= g(\lambda/r^2) \, ,  \qquad 
    g(x)=\left({x}/{b}\right)^{\frac{1}{n}} \, ,
\end{eqnarray}
while the metric coefficient reduces to
\begin{eqnarray}
    f=1 -\frac{2m_0}{\Mpl^2r}+ \frac{\lambda n^2}{\Mpl^2(1+n)(2-n)} \left( \frac{\lambda}{b }\right)^{{1}/{n}} r^{-2/n}-\frac{\Lambda r^2}{3} \, ,
\end{eqnarray}
where $m_0$ is a mass parameter and we have supposed $n \neq 2$ in addition to {$n\neq -1,0$. }
Here $\Lambda$ is the cosmological constant, which we set to zero in the following examples of this subclass. 

In the case where $n=0$, $P$ is a linear function of $X'$ and the equation for $X'$ reduces to $Y=0$ as we can see from \eqref{eqYX}. Hence, either $\psi'=0$ or $X'=0$. In that case, the geometry reduces to a Schwarzschild metric, hence  we get a stealth solution with a constant $X$. 

The case $n=1$ is particularly interesting since the background is described by a Reissner-Nordstrom metric with a charge parameter $Q_{\rm RN}^2=\lambda^2/(2b)$. This is to be expected, since in this case the action is analogous to that of a (usual) linear Maxwell theory coupled to gravity. Furthermore, we recover the expected \(Q_{\rm RN}/r^2\) behavior for \(X'\), which plays the role of the electric field up to a normalization factor.

Even though the case $n=2$ needs to be treated separately, it raises no problem in the computation of the metric coefficient. Indeed, after a direct calculation, one shows that 
\begin{eqnarray}
    f= 1-\frac{2m_0}{\Mpl^2r} - \frac{2\lambda}{3\Mpl^2} \left( \frac{\lambda}{b}\right)^{1/2} \frac{\ln(r/r_{\rm ref})}{r} \, , 
\end{eqnarray}
where $r_{\rm ref}>0$ is a reference length whose change can be absorbed into $m_0$. The metric therefore exhibits a pathological logarithmic behaviour.

\subsection{More general Lagrangians}
\label{sec:Quadratic-theories}

In this section, we consider more general Lagrangians $P(X,{\cal F})$ which depend on ${\cal F}$ (or equivalently on $X'$) but now with a non trivial dependency on $X$, and we want to compute their associated background solutions. 

We therefore study the solutions of  the equation \eqref{eq:SF-master} for $X(r)$ 
which can be trivially reformulated as follows
\begin{eqnarray}
 P_{X' X'} \,  X'' + P_{X'X} \, X' + \frac{2}{r} P_{X'} - P_X \; = \; 0 \, .
\end{eqnarray}
The equation is quasi-linear, i.e. linear in the highest derivative $X''$, and contains friction terms involving the first derivative $X'$, which prevents it from being integrable in full generality. However, there exist cases where the equation is exactly solvable. 

\subsubsection{Up to second order in $X'$}

\label{sec:Non-linear-models}

We first consider the class of theories where $P$ is a second order polynomial in $X'$, i.e. it takes the  form
\begin{eqnarray}
    P = P_0(X) + P_1(X) \, X' + \frac{1}{2} P_2(X) \, X'^2 \, . 
\end{eqnarray}
Assuming $P_2\neq0$, the equation of motion for $X$ becomes
\begin{eqnarray}
    X'' + \frac{2}{r} X' + \Gamma(X) X'^2 + \frac{1}{r} \omega_1(X) + \omega_0(X) = 0 \, ,
\end{eqnarray}
where we have introduced the functions
\begin{eqnarray}
    \Gamma=\frac{1}{2}\frac{P_{2X}}{P_2} \, , \qquad\omega_1= 2\frac{P_1}{P_2} \, , \qquad\omega_0=- \frac{P_{0X}}{P_2} \, .
\end{eqnarray}
Let us show that a simple redefinition of the variable $X$ enables us to remove the non-lineaity in $X'$. Indeed, if we define the new variable $\tilde X$ according to, 
\begin{eqnarray}
\label{refofX}
    \tilde X = G(X) \qquad \text{with} \qquad G_X^2={\vert P_2 \vert } \, . 
\end{eqnarray}
we easily show that the new variable $\tilde X$ satisfies now the equation
\begin{eqnarray}
    \tilde X''+ \frac{2}{r} \tilde X' + \frac{1}{r} \tilde\omega_1 + \tilde\omega_0 = 0 \, ,
\end{eqnarray}
with  no more nonlinearity  in $\tilde X'$ and $\tilde \omega_0$ and $\tilde \omega_1$ are now functions of $\tilde X$ defined by
\begin{eqnarray}
    \tilde\omega_n = G_X \, \omega_n \, ,\quad n = 0,1 \, .
\end{eqnarray}
When $G_X \neq 0$, we can locally invert the relation \eqref{refofX} and express $X$ in terms of $\tilde X$ by virtue of the implicit function theorem\footnote{As an example, let us assume that $\Gamma$ is a nonzero constant, in which case
\begin{eqnarray}
    P_2(X) = p^2 \exp \left( 2 \Gamma X\right) \quad \Rightarrow \quad \tilde X=G(X)=\frac{p}{\Gamma} \exp \left( \Gamma X\right) + \tilde X_0 \, ,
\end{eqnarray}
where we have assumed without loss of generality that $P_2(X)>0$ and $\tilde X_0$ is the integration constant. As a consequence, we can immediately invert this relation
to get $X$ in terms of $\tilde X$:
\begin{eqnarray}
    X = \frac{1}{\Gamma} \ln \left[{\Gamma} (\tilde X-\tilde X_0)/p\right] \, .
\end{eqnarray}
}. The transformed equation has no term quadratic in $\tilde X'$. Hence, without loss of generality, we may assume that $\Gamma$ vanishes. In that case, $P_2$ is a constant. 

For simplicity, we further discard all nonlinearities in the equation, which amounts
to taking $\omega_1$ and $\omega_0$ to be affine functions of $X$:
\begin{eqnarray}
    \omega_n(X) = \alpha_n + \beta_n X \, , 
\end{eqnarray}
where $\alpha_n$ and $\beta_n$ are constant for $n \in \{0,1\}$. 
The general solution of the equation is rather involved, although it admits simpler expressions in some particular cases. We are going to study these cases successively in the following sections. 

\subsubsection{Yukawa branch}

The case where $\omega_1=0$, i.e. $P_1=0$ is particularly interesting. Furthermore, we assume that $P_2=\sigma \in \{-1,+1\}$ and therefore the scalar part of the Lagrangian reduces to 
\begin{eqnarray}
P (X,{\cal F})= \sigma \left[\gamma_0-\left(\alpha_0 X + \frac{1}{2}\beta_0X^2 \right) - \frac{1}{2q^2} {\cal F} \right]\, ,
\end{eqnarray}      
where $\gamma_0$ is a constant parameter of the theory. Following the analysis of the previous subsection, the solution for $X(r)$ does not depend on $\sigma$ and it reads
\begin{eqnarray}\label{eq:Yukawa-solution}
        X(r)= X_0 + \frac{1}{r} \left[ C_1 e^{\mu r} + C_2 e^{-\mu r} \right] \, , \qquad X_0 = \alpha_0/\mu^2\, ,
\end{eqnarray}
where $\mu=\sqrt{-\beta_0}>0$ is the inverse screening length, with $\beta_0<0$ giving exponential rather than oscillatory behaviour\footnote{If $\beta_0$ vanishes, the solution  becomes
    \begin{eqnarray}\label{eq:Constant-force-solution}
        X(r)=-\frac{\alpha_0}{6} r^2 + C_1 + \frac{C_2}{r} \, .
    \end{eqnarray}}.   
	In order to obtain a configuration that remains bounded at large $r$, we focus on the decaying branch $C_1=0$, for which
	\begin{align}\label{eq:X-sol-massive-decay}
		X(r)=X_0+\frac{C_2}{r}e^{-\mu r}.
	\end{align}
	Substituting \eqref{eq:X-sol-massive-decay} into \eqref{massfunction} and integrating, we obtain the metric
\begin{eqnarray}
    f(r) = 1-\frac{2m_0}{\Mpl^2r} + \frac{\sigma Q_\phi^2}{\Mpl^2r^2}e^{-2\mu r} - \frac{\Lambda r^2}{3} \, ,
\end{eqnarray}
where $m_0$ is an integration constant and we have introduced the constants
\begin{eqnarray}
Q_\phi^2=\frac{C_2^2}{2} \, , \qquad \Lambda=\frac{\sigma}{2\Mpl^2}\left(\frac{\alpha_0^2}{\mu^2}-2\gamma_0\right) \, .
\end{eqnarray}
Here $Q_\phi$ is a scalar hair parameter and $\Lambda$ is the effective cosmological constant. The metric diverges as $r^2$ at infinity unless $\Lambda$ vanishes, which corresponds to the case in which $P$ also tends to zero at infinity, as expected. Indeed, we have
\begin{eqnarray}
    \lim_{r \rightarrow + \infty} P(X(r),{\cal F}(r)) = \sigma \left[\gamma_0 - \alpha_0 X_0 - \frac{1}{2} \beta_0 X_0^2 \right] = -\Lambda \Mpl^2\, .
\end{eqnarray}
Thus, we assume this condition to be satisfied from now on and it corresponds to  
\begin{eqnarray}
    P(X,{\cal F})= \sigma\left[\frac{\mu^2}{2}(X-X_0)^2 - \frac{1}{2q^2} {\cal F} \right] \, .
\end{eqnarray}
In that case, the geometry is asymptotically flat in the standard Schwarzschild sense. The deviation from the Schwarzschild solution takes the form of a Yukawa-like term with decay length $1/(2\mu)$, while $X-X_0$ has screening length $1/\mu$.

For $m_0>0$ and nonzero scalar hair, the horizon radii, if there exist any, are solutions of the transcendental equation
\begin{align}
f(r) \equiv 1-\frac{2m_0}{\Mpl^2r}+\frac{\sigma Q_\phi^2}{\Mpl^2r^2}e^{-2\mu r}=0, \label{eqforh}
\end{align}
which in general cannot be solved in closed form. In the case where $\sigma=-1$, there is one  and only one horizon located at $r_h$ since 
    \begin{eqnarray}
       \lim_{r\rightarrow 0} f(r) =- \infty \, , \quad
        \lim_{r\rightarrow +\infty} f(r) = 1 \, 
       \quad \text{and} \quad f'(r) >0  \quad \text{for} \; r >0.
    \end{eqnarray}
One can obtain an approximate expression of $r_h$ in  some limiting cases. For instance, when $\mu m_0/\Mpl^2 \gg 1$, we can neglect the exponential at the leading order and, at the sub-leading order, we have
\begin{eqnarray}
r_h \; \simeq \; \frac{2m_0}{\Mpl^2} + \frac{Q_\phi^2}{2m_0} e^{-4\mu m_0/\Mpl^2} \, .
\end{eqnarray}
In the case where $\sigma=+1$, the resolution of \eqref{eqforh} is  subtler and it is more convenient to study the roots of $H(r)=r^2 f(r)$ which satisfies 
    \begin{eqnarray}
&& H(0)=\frac{Q_\phi^2}{\Mpl^2}>0 \, , \qquad  \lim_{r \rightarrow +\infty} H(r) = +\infty \, , \\
&&H'(0)= -\frac{2(m_0+\mu Q_\phi^2)}{\Mpl^2}<0 \, , \qquad H''(r)>0 \, .
\end{eqnarray}
Therefore, $H(r)$ decreases up to  a minimum and then increases to infinity. Thus, $H(r)=0$ admits zero, one or two solutions, which implies that the metric has no, one or two horizons depending on the parameters. If the metric admits a single horizon $r_h$, this horizon must be a double root $r_c$ of $H$, i.e. it must satisfy simultaneously $H(r_c)=0$ and $H'(r_c)=0$, which leads to
\begin{equation}
\begin{aligned}
 r_c = \frac{1}{2\mu}\left[\frac{2\mu m_0}{\Mpl^2}-1
       +\sqrt{1+\frac{4\mu^2m_0^2}{\Mpl^4}}\right] \, ,\qquad
 Q_{\phi,c}^2 = \frac{\Mpl^2r_c-m_0}{\mu}e^{2\mu r_c} \, .
\end{aligned}
\end{equation}
As a consequence, when $Q_\phi^2 > Q_{\phi,c}^2$, the equation has no solutions and the metric describes a naked singularity. When $Q_\phi^2 < Q_{\phi,c}^2$, the equation admits two distinct solutions and the metric admits two horizons, very similarly to the charged black hole in GR. Hence, $Q_\phi$ controls the scalar hair contribution to the metric and the case where $Q_\phi^2=Q_{\phi,c}^2$ corresponds to the extremal configuration. 

Let us quickly comment on the scalar field profile.
As we showed in \eqref{eq:X-sol-massive-decay}, the invariant $X(r)$ approaches $X_0$ exponentially fast as $r\to\infty$.  From $\psi'^2=(X+q^2/f)/f$, one finds asymptotically $\psi'^2\to X_0+q^2$, so the radial scalar gradient itself vanishes at infinity only for the special choice $X_0=-q^2$.

To conclude, let us emphasize that the spacetime approaches the Schwarzschild geometry at large distances, while the nonlinear scalar hair modifies the geometry only at finite radii because of the Yukawa falloff.

\subsubsection{Other branches}    
The analysis of the previous subsection can be extended to many other cases in which the equation for $X(r)$ remains linear (possibly with a constant source).

In the case where $\omega_0=0$, i.e. $P_{0X}=0$, the 
theory if of the form
\begin{eqnarray}
P(X,{\cal F})= P_0 + P_2 \left[\frac{1}{2q}  (\alpha_1 + \beta_1 X) (-{\cal F})^{1/2} - \frac{1}{2q^2} {\cal F} \right] \, ,
\end{eqnarray}
with $P_0$ and $P_2$  constant, and the    solution for $X(r)$ reads
\begin{eqnarray}
X(r)= -\frac{\alpha_1}{\beta_1} + {\frac{1}{x} }\left[ C_1 J_1\left(x\right) + C_2 Y_1\left(x\right)\right] \, , \qquad x \equiv 2 \sqrt{r/r_1} \, ,
\end{eqnarray}
where $\beta_1 r_1=1$ and the functions $J_\nu$ and $Y_\nu$ are Bessel functions of the order $\nu$ of first and second kind, respectively. Here, we also assume $\beta_1 \neq 0$. If this is not the case, the solution becomes  
\begin{eqnarray}
X(r)=-\frac{\alpha_1}{2} r + C_1 + \frac{C_2}{r} \, .
\end{eqnarray}

In the case where the equation is linear (with no source terms), i.e. $\alpha_0=\alpha_1=0$, the theory is of the form
\begin{eqnarray}
P(X,{\cal F})= \gamma_0 + P_2 \left[ - \frac{\beta_0}{2} X^2 + \frac{\beta_1}{2q} X (-{\cal F})^{1/2} - \frac{1}{2q^2} {\cal F} \right] \, ,
\end{eqnarray}
with $\gamma_0$ and $P_2$ constants, and the solution for $X(r)$ reads:
\begin{eqnarray}
X(r)= \frac{1}{r}\left[ C_1 M_{\kappa,1/2}(2 r/r_0) + C_2 W_{\kappa,1/2} (2 r/r_0)\right] \, ,
\end{eqnarray}
where $\beta_0 r_0^2=-1$ and $2\kappa= \beta_1 r_0$. We are using the standard notations $W_{\kappa,\nu}$ and $M_{\kappa,\nu}$ for the Whittaker functions and we have assumed that $\beta_0 \neq 0$. 
    
Finally, the case where $\beta_0=\beta_1=0$ is rather simple  to study. It corresponds to a theory  of the form
   \begin{eqnarray}
   \label{constantforce}
       P(X,{\cal F}) = \gamma_0 + P_2 \left[ - \alpha_0 X + \frac{\alpha_1}{2q} (-{\cal F})^{1/2} - \frac{1}{2q^2} {\cal F} \right] \, ,
   \end{eqnarray}
with $\gamma_0$ and $P_2$ constant, and the solution reads:
    \begin{eqnarray}
        X(r)=-\frac{\alpha_0}{6} r^2 - \frac{\alpha_1}{2}r +C_1+\frac{C_2}{r} \, .
    \end{eqnarray}
In all these solutions, $C_1$ and $C_2$ are integration constants.

On can compute the metric in all these cases (at least as an integral expression for $f(r)$). However, for illustration, we only consider the last case associated to the theory \eqref{constantforce} where the equation for $f(r)$ can be solved explicitly to get
\begin{eqnarray}
    f(r) = 1-\frac{2m_0}{\Mpl^2r}-\frac{\Lambda r^2}{3}
    +\frac{1}{\Mpl^2}\left[\frac{a_{-2}}{r^2}+a_0+a_1r+a_3r^3+a_4r^4\right]\, ,
\end{eqnarray}
where $m_0$ is an integration constant and
\begin{equation}
 \Lambda\equiv\frac{-\gamma_0+P_2C_1\alpha_0+P_2\alpha_1^2/8}{\Mpl^2}\,.
\end{equation}
The remaining coefficients are given by
\begin{eqnarray*}
    a_{-2}= \frac{P_2 C_2^2 }{2}\, , \quad
    a_0 = -\frac{P_2C_2\alpha_1}{2} \, , \quad
    a_1= -\frac{2P_2 C_2 \alpha_0}{3} \, , \quad
    a_3= \frac{P_2 \alpha_0 \alpha_1}{12} \, , \quad
    a_4= \frac{P_2\alpha_0^2}{45} \, .
\end{eqnarray*}
The conditions for the space-time to be (locally) asymptotically flat are $\alpha_0=0$ and $\Lambda=0$, the latter giving $\gamma_0=P_2\alpha_1^2/8$. These conditions set $a_1=a_3=a_4=0$, but $a_0$ is generally nonzero, which means that the metric has a deficit angle. To remove this angle, one needs to impose the extra condition $\alpha_1 C_2 =0$. If $C_2=0$, the metric reduces to Schwarzschild and we have a stealth solution. The case $\alpha_1=0$ is more interesting since the metric has a Reissner-Nordstr\"om form with a charge parameter $Q_{\rm RN}$ given by $Q_{\rm RN}^2=a_{-2}$. Notice that $Q_{\rm RN}^2$ can be negative in the case where $P_2<0$. 

\subsection{Nonlinear models}
In this section, we consider theories defined by a Lagrangian that  factorizes as
    \begin{eqnarray}
   P(X,{\cal F})=Q(X) {\cal F}  \,,
   \end{eqnarray}
   where $Q$ is an arbitrary function of $X$. Interestingly the equation for $X$ \eqref{eq:SF-master} admits a first integral since it can be shown to be equivalent to
    \begin{eqnarray}
        \frac{\D}{\D r} \ln \left\vert r^4  Q(X) {\cal F}\right\vert \; = \; 0 \, .
    \end{eqnarray}
As a consequence, $r^4 X'^2 Q(X)=C_1$ where $C_1$ is a constant fixed by initial conditions and $Q$ has a fixed  sign (given by the sign of the constant $C_1$). This last equation can in turn be integrated into 
\begin{eqnarray}
\int \D X \sqrt{Q(X)/C_1}  = -\frac{\operatorname{sign}(X')}{r} + C_2 \, ,
\end{eqnarray}
which enables us to express (at least locally) $X$ as a function of $r$. Here $C_2$ is a new integration constant.

The expression of the metric follows immediately from \eqref{massfunction}:
\begin{equation}
    f(r)=1-\frac{2m_0}{\Mpl^2r}-\frac{1}{\Mpl^2r} \int^r \D x \, x^2 Q(X){\cal F}  \; = \; 1 - \frac{2m_0}{\Mpl^2r}-\frac{q^2 C_1}{\Mpl^2 r^2} \, ,
\end{equation}
where $m_0$ is again an integration constant. This metric coefficient is independent of the detailed form of $Q(X)$ and reduces to a Reissner-Nordström metric with charge parameter $Q_{\rm RN}^2=-q^2C_1$. The charge is real when $C_1<0$ and purely imaginary when $C_1>0$. The case $C_1=0$ reduces to the Schwarzschild solution, which is expected since in the case $X' \neq 0$ it corresponds to $Q=0$ and the theory reduces to Einstein-Hilbert.

\subsection{The special family}
\label{sec:exceptional-background}

To finish this section on analytic solutions, it is interesting for what follows to consider also the special family of theories\footnote{This class corresponds to $P_0=K$, $P_1=\sigma_XqQ$ and $P_2=0$ in Sec.~\ref{sec:Non-linear-models}, and therefore requires a separate treatment. }
\begin{equation}
 P(X,{\cal F})=K(X)+Q(X)\sqrt{-{\cal F}} \,,
 \label{eq:exceptional-P}
\end{equation}
where $K$ and $Q$ are arbitrary functions of $X$. Its ``special'' role  will be explained in Subsec.~\ref{sec:exceptional} when we study the monopole perturbations. Notice that the interaction $Q(X)\sqrt{-{\cal F}}$ is non-polynomial in $\phi_{\mu\nu}$ and lies outside the quadratic UDHOST action considered in \cite{DeFelice:2022xvq}. 

To solve the equations of motion, we assume that
$q>0$ and consider a static interval with $f>0$, $X<0$ and $\psi'X'Q\neq0$.
Substituting~\eqref{eq:exceptional-P} into \eqref{eq:metric-master} and \eqref{eq:SF-master} gives the following two equations\footnote{The $Q$-dependent contribution to the radial Lagrangian for~\eqref{eq:SF-master} is $r^2\sigma_X qQ(X)X'$. Although linear in $X'$, it is not a boundary term because of the factor $r^2$. Integration by parts leaves a bulk term with explicit radial dependence, whose variation gives the right-hand side of the first equation in~\eqref{eq:exceptional-bg}. Hence the setup is not equivalent to a covariant k-essence theory. } for the scalar field and the metric
\begin{equation}
	K_X=\frac{2\sigma_X qQ}{r}\,,\qquad
	\Mpl^2[r(f-1)]'=r^2K\,,
	\label{eq:exceptional-bg}
\end{equation}
where we used $f=h$ and the relations
\begin{eqnarray}
    \sqrt{-{\cal F}}=\sigma_X qX' \, , \qquad P_{\cal F}=-\frac{\sigma_X Q}{2qX'} \, ,\qquad \sigma_X\equiv\operatorname{sign}(X')=\pm1 \, .
\end{eqnarray}
We assume $\sigma_X$ to be constant on the interval we are considering here. 
For given $K$ and $Q$, the first equation in~\eqref{eq:exceptional-bg} determines $X(r)$, while the second gives the metric by integration, as in~\eqref{massfunction}. Thus the general local construction on this branch is
\begin{equation}
	r=\frac{2\sigma_X qQ(X)}{K_X(X)}\,,\qquad
	f(r)=1-\frac{2m_0}{\Mpl^2r}+\frac{1}{\Mpl^2r}\int^r x^2K(X(x))\,\D x\,.
	\label{eq:exceptional-construction}
\end{equation}
The first relation must be invertible to give a differentiable profile $X(r)$. In the second relation, the additive constant in the integral is absorbed into $m_0$. In the very peculiar case where $Q=0$, the first equation in~\eqref{eq:exceptional-bg} instead gives $K_X=0$, as discussed further in Subsec.~\ref{sec:exceptional}. In order to compute the scalar field $\psi(r)$, we recall that \eqref{eq:psi-reconstruction}  requires the condition $q^2+fX>0$ for a real nonzero $\psi'$. 

Let us now investigate a concrete example of black hole solution for the choice~\eqref{eq:exceptional-P} of function $P$. We take the functions $K$ and $Q$ to be monomials of $-X$,
\begin{equation}
    K=-c\,(-X)^{\alpha}\,,\qquad Q=Q_0\,(-X)^{\beta}\,,\qquad c,\,Q_0,\,\alpha>0\,,
\end{equation}
which gives, after solving~\eqref{eq:exceptional-construction},
\begin{equation}
\begin{aligned}
 &X=-\frac{\lambda}{r^{n}}\,,\qquad n=\frac{1}{\alpha-\beta-1}\,,\qquad
 \lambda=\Big(\frac{2qQ_0}{c\,\alpha}\Big)^{n}\,,\\
 & f=1-\frac{2m_0}{M_{\rm Pl}^2r}+\frac{c\,\lambda^{\alpha}}{M_{\rm Pl}^2(n\alpha-3)}\,r^{2-n\alpha}\,.
\end{aligned}
\label{eq:BH-special-branch}
\end{equation}
One can restrict the possible values of $\alpha$ and $\beta$ using the monopole stability criteria obtained later in~\eqref{eq:exceptional-H-positivity}. In this situation one has two conditions. The first one $Q > 0$ is satisfied by construction since $Q_0>0$. The second one  $(rX)'=(n-1)\lambda r^{-n}>0$ requires $n>1$, i.e.
\begin{equation}
    \alpha<\beta+2 \,.
\end{equation}
Notice that  we also require the condition $n>0$, which translates into $\alpha>\beta +1$, so that $X$ vanishes at infinity. 
Asymptotic flatness with a finite mass further requires $2\alpha<3(\beta+1)$, which implies $\beta>-1$.

Choosing $\alpha = 2$, we can take $\beta = 1/2$, and the function $P$ along with $X$ and the black hole line element are described by
\begin{eqnarray}
 P=-cX^2+Q_0\sqrt{X{\cal F}}\,,\qquad
 X=-\frac{q^2Q_0^2}{c^2r^2}\,, \qquad
 f=1-\frac{2m_0}{M_{\rm Pl}^2r}+\frac{q^4Q_0^4}{c^3M_{\rm Pl}^2r^2}\,.
\end{eqnarray}
More generally, for $3\alpha=4(\beta+1)$ with $-1<\beta<2$, the metric is exactly Reissner-Nordström, with $Q_{\rm RN}^2=c\lambda^\alpha/M_{\rm Pl}^2>0$ as in the previous example.



\section{Perturbations}
\label{sec:perturbations}
We now study axial and monopole perturbations of the background solutions in F-UDHOST gravity. Throughout this section, we use units in which $\Mpl=1$.

\subsection{Axial perturbations}
The dynamics of axial perturbations around static and spherically symmetric solutions are relatively simple to study, due to two complementary reasons. First, the tensor modes are entirely encoded in the Ricci scalar exactly as in GR. Second, unlike in Einstein gravity coupled to NLED, the scalar sector does not introduce an additional axial degree of freedom. As a consequence, there is a single axial degree of freedom whose dynamics are very similar to that of the axial degree of freedom in Einstein gravity.

To be concrete, we start by considering a metric perturbation 
\begin{align}\label{eq:metric-pert}
h_{\mu\nu}\equiv\delta g_{\mu\nu}= g_{\mu\nu} - \overline{g}_{\mu\nu} \,,
\end{align}
around the background metric $\overline{g}_{\mu\nu}$. The scalar perturbation is not involved in the axial sector hence it vanishes identically, i.e. $\delta \phi= \phi - \overline{\phi}=0$ with $\overline{\phi}$ the background scalar field.   Following the traditional Regge-Wheeler analysis~\cite{Regge:1957td}, axial metric perturbations can be expanded in spherical harmonics $Y_{\ell m}(\theta,\varphi)$ and, after a gauge fixing, the non-trivial component of $\delta g_{\mu\nu}$ can be reduced into the following form:
\begin{align}
	 & h_{t\theta} = \frac{1}{\sin\theta}  \sum_{\ell, m} h_0^{\ell m}(t,r) \partial_{\varphi} {Y_{\ell m}}(\theta,\varphi), \qquad
	h_{t\varphi} = - \sin\theta  \sum_{\ell, m} h_0^{\ell m}(t,r) \partial_{\theta} {Y_{\ell m}}(\theta,\varphi), \nonumber         \\
	 & h_{r\theta} =  \frac{1}{\sin\theta}  \sum_{\ell, m} h_1^{\ell m}(t,r)\partial_{\varphi}{Y_{\ell m}}(\theta,\varphi), \qquad
	h_{r\varphi} = - \sin\theta \sum_{\ell, m} h_1^{\ell m}(t,r)  \partial_\theta {Y_{\ell m}}(\theta,\varphi) \, .\label{eq:odd-perttext}
\end{align}
We only consider modes for $\ell \geq 2$ since there is no monopole ($\ell =0$) or radiative dipole ($\ell=1$) in axial perturbations. Furthermore, at the linear level, modes with different angular momenta $\ell$ do not couple to one another. We will therefore omit the  indices $\ell$  and $m$ from the functions $h_0$ and $h_1$.

The next step consists in computing the equations of motion satisfied by the functions $h_0$ and $h_1$. We follow the method developed in \cite{Langlois:2021aji,Langlois:2022ulw} in the context of DHOST theories \cite{Langlois:2015cwa} which is based on a linearization of equations of motion \eqref{Einsteinequationstensortext}. One can equivalently start with the calculation of the quadratic action as it was initiated and developed in \cite{Kobayashi:2012kh,Kobayashi:2014wsa} still in the context of higher-derivative scalar-tensor theories (see also \cite{Aoki:2021wew} for Effective Field Theory approach). These approaches are obviously equivalent. 

It is straightforward to see that the perturbation $\delta X = X - \overline{X}$ vanishes in the axial sector, i.e.
\begin{eqnarray}
    \delta X = \delta g^{\mu\nu} \, \overline{\phi}_\mu \overline{\phi}_\nu = -\overline{g}^{\mu\alpha}\overline{g}^{\nu\beta}h_{\alpha\beta} \, \overline{\phi}_\mu \overline{\phi}_\nu = 0 \, .
\end{eqnarray}
Similarly, $\delta {\cal F}=0$ and therefore, in the theory we are considering here, the perturbed equations reduce to
\begin{eqnarray}
    \delta G_{\mu\nu} - P \, \delta g_{\mu\nu} \; = \; 0 \, ,
\end{eqnarray}
where $P$ is evaluated on the background solution as well. From now on, for simplicity, we will drop the bars on all background quantities when there is no possible confusion.
Next, we  decompose the two functions $h_0$ and $h_1$ in Fourier modes:
\begin{eqnarray}
\label{Fourtrans}
    h_a(t,r) = \int \D \omega \, e^{-i \omega t} \, h_a(\omega,r)  \, , \quad \text{for} \; a=0,1 \,,
\end{eqnarray}
using the same notation for $h_a$  in the time and frequency domains for simplicity. 
Then, following exactly the same method as in \cite{Langlois:2021aji}, we show that there are only two independent components for the axial perturbations equations which are
given by
\begin{eqnarray}
    &&\omega h_0 - i f \left( f h_1  \right)' = 0 \, , \\&& h_0' - \frac{2h_0}{r} + i\omega h_1 - \frac{i f}{\omega r^2}(2 \lambda_\ell + 2rf' + r^2 f'' - 2r^2 P)h_1 =0 \, ,
\end{eqnarray}
where $\lambda_\ell= (\ell^2+\ell -2)/2$. As shown in Appendix \ref{App:perturbations}, this system reproduces the dynamics of a massless spin 2 field in the background metric. In fact, such a result extends a property previously obtained in DHOST theories \cite{Langlois:2022ulw}, where the perturbation equation coincides with that governing a free massless spin-2 field propagating on an effective metric.
In the context  of this article, the effective metric is the background metric since the original action is written in an Einstein-like frame (where there is no non-minimal coupling to the Ricci scalar). 

To close this analysis, let us recall the very well-known fact that this system of two first order equations is equivalent to a single Schrödinger-like equation 
\begin{eqnarray}
    - \frac{\D^2 \Psi}{\D r_*^2} + V(r) \Psi = \omega^2 \, \Psi \, ,
\end{eqnarray}
where the ``tortoise-like'' coordinate $r_*$, the ``wave function'' $\Psi$ and the effective potential $V$ are defined by
\begin{eqnarray}
\label{effectivemetric}
    \D r_*=\frac{\D r}{f(r)} \, , \quad
    \Psi= \frac{f h_1 }{r}  \, , \quad  V = \frac{f}{r^2} \left( 2\lambda_\ell + 2 f -r f'\right)\, .
\end{eqnarray}
This decomposition allows us to immediately recognize a S-deformation \cite{Kimura:2017uor}:
\begin{equation}
V=S^2-\partial_* S+V_\lambda, \qquad V_\lambda=2 \lambda_\ell \frac{f}{r^2}, \qquad S \equiv \frac{1}{2} \frac{\partial_* r^2}{r^2},
\end{equation}
As a consequence, axial perturbations are stable when the background geometry describes a static and spherically symmetric, asymptotically flat black hole. In the case where the background is not a black hole (it could be a naked singularity or a soliton, for instance), the stability analysis requires further investigation as it was shown in \cite{Noui:2023ksf}. 

\subsection{Monopole perturbations: a no-go result}
\label{sec:Monopole}
The full computation of polar perturbations around static and spherically symmetric solutions is much more involved in scalar-tensor theories than in General Relativity, because of the additional degree of freedom provided by the scalar field. Indeed, in General Relativity, the decoupling between the two modes at the linear perturbation level is made possible by a separation by parity: one obtains two decoupled equations, one describing axial perturbations (the Regge-Wheeler equation~\cite{Regge:1957td}), and the other describing polar perturbations (the Zerilli equation~\cite{Zerilli:1970se}). However, the scalar degree of freedom, being of even-parity,  couples to the polar degree of freedom from the metric, yielding a dynamical system that cannot be explicitly decoupled in general~\cite{Langlois:2021aji}. Therefore, before going to the full study of polar perturbations, we first turn to the monopole perturbations, following~\cite{Charmousis:2026bhe}. In this case, the polar degree of freedom is non-dynamical, and the linear perturbations should boil down to the study of one dynamical mode.


\subsubsection{Quadratic action for the monopole}
Concretely, we proceed similarly to the previous section, considering metric perturbation \eqref{eq:metric-pert} and a scalar perturbation $\delta\phi = \phi - \bar{\phi}$. For monopolar perturbations, the metric and scalar perturbations can be parametrized by
\begin{subequations}
\begin{align}
h_{tt} &= f(r) H_0(t,r) \,,& h_{tr} &= H_1(t,r)\,,& h_{rr} &= \frac{H_2(t,r)}{f(r)} \,,\\
h_{ab} &= K_0(t,r) \bar{g}_{ab} \,,& \delta\phi &= \delta\phi(t,r)\,,
\end{align}
\label{eq:perturbations-parametrization}
\end{subequations}
where latin indices span angular variables. For $q \neq 0$, a gauge choice allows us to set $K_0 = 0$ along with $\delta\phi = 0$, and this choice fixes the gauge completely. We define $S_\text{quad}$ as the coefficient of $\epsilon^2$ in the expansion of the action~\eqref{eq:action}, with $g_{\mu\nu}=\bar g_{\mu\nu}+\epsilon h_{\mu\nu}$. The quadratic order then yields a complex expression $S_\text{quad}$ that contains  quadratic combinations of $H_0$, $H_1$, $H_2$ and their time and space derivatives. However, as we shall see, this expression can be greatly simplified by recognizing combinations of these fields.

First, we introduce
\begin{equation}
    \mathcal{Y} = - \delta X = \frac{q^2}{f} H_0 - 2 q \psi' H_1 + f \psi'^2 H_2  \,.
\end{equation}
Then, after imposing the background equations of motion and integrating by parts, the quadratic action takes the form
\begin{equation}
\begin{aligned}[t]
    S_\text{quad}[\mathcal{Y}, H_1, H_2] = \int \D t\,\D r\,\D \Omega\, \Big[ &- r^2 \mathcal{A} \big(\psi' \partial_t \mathcal{Y} - q \partial_r\mathcal{Y} + c_1 H_1 + c_2 H_2\big)^2 \\
    &+ \frac{rf}{2q^2} H_2 \Big(f \partial_r \mathcal{Y} + 2 q f \psi' \partial_r H_1 - \frac{2q^2}{f} \partial_t H_1\Big) \\
    &+ \frac{r f}{q} (f \psi')' H_1 H_2 + \frac{rff'}{2q^2} H_2 \mathcal{Y} \\
    &+ \frac{r}{q}\, c_3\, \mathcal{Y} \Big(f \psi' H_1 - \frac{2q^2 + fX}{2q} H_2 \Big) + \frac{c_4}{4} H_2^2 + \frac{c_5}{2} \mathcal{Y}^2\Big] \,,
\end{aligned}
\label{eq:Monopole_quad_action}
\end{equation}
where $\D \Omega = \sin\theta\,\D \theta\,\D \varphi$, all the $c_i$ are radial coefficient functions defined in Appendix~\ref{app:Quad_Action_Monopole}, and
\begin{equation}
    \mathcal{A} = P_{\cal F} + 2{\cal F} P_{{\cal F}{\cal F}}\, .
    \label{CalA}
\end{equation}
The stability analysis in this subsection concerns the generic case $\mathcal A\neq0$, with $\mathcal A$ defined above. The special case $\mathcal A=0$ corresponds to theories in the special family~\ref{sec:exceptional-background} and is studied separately in Subsec.~\ref{sec:exceptional}.
Here by $\mathcal{A} = 0$, we mean that $\mathcal{A}$ vanishes identically off shell. 
Throughout this section we also assume $X' \neq 0$ and $X<0$, so that the hypersurfaces of constant $\phi$ are spacelike.

\subsubsection{Simplication of the quadratic action: new variables and new coordinates}

The shape of this quadratic action is quite simple: beyond the complicated mass terms, the derivatives appear in specific combinations. In particular, it is interesting to introduce the scalar time $\tau = t + \psi/q$, such that $\phi = q \tau$, and to use $(\tau, r)$ as coordinates. In the following, $\partial_r$ denotes the radial derivative at fixed $\tau$, related to the one at fixed $t$ by
\begin{equation}
    \left.\frac{\partial}{\partial r}\right|_t = \left.\frac{\partial}{\partial r}\right|_\tau + \frac{\psi'}{q} \frac{\partial}{\partial \tau} \,.
    \label{eq:change-coords-derivatives}
\end{equation}
The Jacobian of the change of coordinates is unity. Replacing $H_1$ by
\begin{eqnarray}
    \Phi = 2q X H_1 + f \psi' \mathcal{Y} \,,
\end{eqnarray}
the quadratic action becomes
\begin{equation}
\begin{aligned}[t]
    S_\text{quad}[\mathcal{Y}, \Phi, H_2] = \int \D \tau\,\D r\,\D \Omega\, \Big[ &-r^2 \mathcal{A} \mathcal{W}^2 + \frac{r f}{2q^2} H_2 \Big[\frac{1}{q} \partial_\tau \Phi + \partial_r\Big(\frac{f \psi' \Phi - q^2 \mathcal{Y}}{X}\Big)\Big] \\
    &+ \frac{c_4}{4} H_2^2 - \frac{r c_3}{2q^2} \mathcal{Y} \Big[(2q^2 + f X) H_2 - \frac{f\psi'}{X} \Phi\Big] + (-X)d_7 \mathcal{Y}^2\Big] \,,
\end{aligned}
\label{eq:quadratic-action-monopole-Phi}
\end{equation}
where the radial coefficient $d_7$ is defined in Appendix~\ref{app:Quad_Action_Monopole}, and we have introduced
\begin{equation}\nonumber
    \mathcal{W} = -q \partial_r \mathcal{Y} + \frac{f \psi' X'}{2qX} (\Phi - f\psi' \mathcal{Y}) + c_2 H_2 \,.
\end{equation}
Choosing $q>0$, we introduce the orthonormal {\it temporal} and {\it spatial} derivatives
\begin{equation}
\begin{aligned}
 D_n &\equiv n^\alpha\partial_\alpha
      =-\frac{1}{\sqrt{-X}}
        \left(\frac Xq\partial_\tau+f\psi'\partial_r\right) \, ,\\
 D_s &\equiv s^\alpha\partial_\alpha
      =\frac q{\sqrt{-X}}\partial_r \, .
\end{aligned}
 \label{eq:exceptional-frame}
\end{equation}
The first derivative follows the future-directed unit normal to the constant-$\phi$ hypersurfaces, while the second is the orthogonal outward radial derivative. Primes continue to denote ordinary radial derivatives of background quantities.

Since the field $H_2$ appears only algebraically, we can integrate it out of the action~\eqref{eq:quadratic-action-monopole-Phi}. Assuming $d_0 \neq 0$, we obtain
\begin{equation}
\begin{aligned}[t]
 S_\text{quad}[\mathcal Y,\Phi]
 =-\int\D\tau\,\D r\,\D\Omega\,(-X)\,\Big\{&
 \frac{1}{4d_0}
 \Big[\frac{rf}{2q^2}(D_n\Phi+d_1\Phi)
       +d_2\mathcal Y+d_3D_s\mathcal Y\Big]^2\\
 &+r^2\mathcal A(D_s\mathcal Y+d_4\mathcal Y-d_5\Phi)^2
 -d_6\Phi\mathcal Y-d_7\mathcal Y^2\Big\} \, .
\end{aligned}
\label{eq:quadratic-action-monopole-reduced}
\end{equation}
where the $d_i$ coefficients are defined in Appendix~\ref{app:Quad_Action_Monopole}.

Given the simple shape of the action~\eqref{eq:quadratic-action-monopole-reduced}, we can understand the dynamics of the monopole perturbations in the present context. First, we note that $\mathcal{Y}$ is not a dynamical field: $\Phi$ is the only dynamical field, while $\mathcal{Y}$ is fixed on each time slice. This mode is therefore the counterpart of the \textit{shadowy} mode from UDHOST theories~\cite{DeFelice:2018ewo}. It is then possible to study the propagation of $\Phi$ to ensure this degree of freedom is healthy.

\subsubsection{Short wavelength approximation}

To study the stability of the monopole, we consider local radial Fourier modes
\begin{equation}
 \big(\Phi,\mathcal Y\big)
 \sim\big(\Phi_k(\tau),\mathcal Y_k(\tau)\big)e^{ikr}
 \,,\qquad \partial_r\to ik \, ,
\end{equation}
where $k>0$ is the radial coordinate wavenumber at fixed $\tau$, with arbitrary time dependence of the amplitudes. We require\footnote{The high wavenumber expansion is taken at each regular exterior point, rather than as a uniform limit at the horizon. If the theory is used as an effective field theory, the wavelengths considered must also lie within its range of validity.}
\begin{equation}
 k\gg\max\left\{\frac1r,\left|\frac{f'}f\right|,
                         \left|\frac{X'}X\right|\right\} \, ,
\end{equation}
with analogous inequalities for the other background scales. Considering that the fields $\Phi$ and $\mathcal{Y}$ are of order 1, their radial derivatives are of order $k$, and the reduced quadratic action~\eqref{eq:quadratic-action-monopole-reduced} becomes at its highest order $k^2$
\begin{equation}
\begin{aligned}[t]
 S_\text{quad}\simeq-\int\D\tau\,\D r\,\D\Omega\,(-X)\,\Big\{&
 \frac{r^2f^2}{16q^4d_0}
 \Big[D_n\Phi+\frac qf
       \big(f+2r\mathcal AX(2q^2+fX)X'\big)D_s\mathcal Y\Big]^2\\
 &+r^2\mathcal A(D_s\mathcal Y)^2\Big\} \, .
\end{aligned}
\end{equation}
The Euler-Lagrange equation for $\mathcal Y$ can be integrated once. The integration constant vanishes for perturbations localised in $r$, giving
\begin{equation}
D_s\mathcal Y=-\frac fq
\frac{f+2r\mathcal AX(2q^2+fX)X'}{\Delta}D_n\Phi \, .
\end{equation}
with
\begin{equation}
\label{Delta}
\Delta \equiv f^2 + 4\mathcal{A}X \big[2q^2 (rfX)' + f^2 X^2\big] \,.
\end{equation}
Injecting this solution back into $S_\text{quad}$ finally gives us
\begin{equation}
 S_\text{quad}\simeq-\int\D\tau\,\D r\,\D\Omega\,(-X)\,
 \frac{r^2f^2\mathcal A}{q^2\Delta}(D_n\Phi)^2 \, .
 \label{eq:high-k-action}
\end{equation}
Note that $d_0$ has dropped out, hence this result only requires the condition $\Delta \neq 0$. For $\Delta = 0$ the above reduction breaks down and one has to make the analysis in this specific case. For $\mathcal{A} = 0$, the displayed leading term vanishes and the constraints must be reduced separately, as in the subsection~\ref{sec:exceptional}.

In the following, we assume that $\mathcal A$ and $\Delta$ remain finite and nonzero throughout the connected static exterior. Furthermore, the outer horizon $r_h$ is supposed to be regular and non extremal with $f(r_h)=0$, $f'(r_h)>0$ and $\mathcal A(r_h)\neq0$. We also recall that the conditions $q\psi'X'\neq0$ and $X<0$ are still valid.
With these conditions, we can expand the regular background near the horizon as $f(r)=f'_h(r-r_h)+\mathcal O[(r-r_h)^2]$ where 
$f'_h=f'(r_h)$  and we can show that the quadratic
action~\eqref{eq:high-k-action} becomes
\begin{equation}
 S_\text{quad}\simeq-\int\D\tau\,\D r\,\D\Omega\,(-X)\,
 \Big[\frac{r_hf'_h}{8q^4X_h^2}(r-r_h)^2(D_n\Phi)^2
      +\mathcal O\big[(r-r_h)^3\big]\Big] \, ,
\end{equation}
with $X_h=X(r_h)$, independently of $\mathcal A$. As a consequence, the  kinetic sign is negative
immediately outside the horizon. Moreover, since $\mathcal{A}/\Delta$ is continuous and nonzero and 
\begin{equation}
 \left.\frac{\mathcal A}{\Delta}\right|_{r_h}
 = \frac{1}{8q^2r_h f'_h X_h^2} > 0\,,
 \label{eq:monopole-horizon-ratio}
\end{equation}
the kinetic sign remains negative throughout the connected static exterior. This shows an instability of the monopole and a Hamiltonian counterpart of this result is presented in Appendix \ref{sec:monopole-hamiltonian-generic} where we show that the Hamiltonian is unbounded from below. 

We do not call this negative sign a ghost, because in the present model one cannot use
the sign of the principal part to infer the presence of a ghost in a variable-independent
way. Indeed, one can use for instance the variable 
\begin{equation}
 \hat\Phi \equiv \Phi - \frac{X(2q^2+fX)}{f\psi'}\,H_2 \, .
 \label{eq:Phi-hat}
\end{equation}
This variable is connected with $\Phi$ by an invertible canonical transformation in the Hamiltonian formulation. Next we obtain the reduced quadratic action in terms of variable $\hat{\Phi}$ and only after proceed with the high-$k$ expansion.  After that we arrive at the principal part
\begin{equation}
 S_\text{quad}\simeq-\int\D\tau\,\D r\,\D\Omega\,(-X)\,
 \frac{r^2\mathcal A}{q^2}\,(D_n\hat\Phi)^2 \, . 
\end{equation}
Its sign is positive for $\mathcal A<0$. The sign can change because $H_2$ is
proportional to the momentum conjugate to $\Phi$ and the mode has no gradient term. 

The unbounded quadratic Hamiltonian and the absence of an independent spatial gradient term raise concerns about instability and strong coupling~\cite{Motohashi:2019ymr,Takahashi:2021bml}, respectively. Notice that, even if the classical solution does not exhibit exponential growth, this does not eliminate the negative-energy problem in the interacting theory~\cite{Ganz:2020skf}. 

\subsection{The special branch}
\label{sec:exceptional}

The case $\mathcal A=0$ requires a separate analysis which is presented in this section. After analyzing the stability of the system from the Hamiltonian perspective, we turn to some interesting properties of its propagating degree of freedom.

\subsubsection{Stability analysis}
Let us start with the trivial case of k-essence for which $P_{\cal F}=0$. On the branch $q\psi'X'\neq0$ with $X<0$,~\eqref{eq:SF-master} reduces to $P_X=0$. Thus, the system admits a non-trivial solution only if $P$ is constant (over the range of $X$ covered by the background). In that case,~\eqref{eq:metric-master} gives the Schwarzschild solution with a cosmological constant. The scalar degree of freedom can be shown\footnote{The coefficient definitions in Appendix~\ref{app:Quad_Action_Monopole} then give $c_3=c_5=d_7=0$. Substituting these relations and $\mathcal A=0$ into~\eqref{eq:quadratic-action-monopole-Phi}, variation with respect to $\mathcal Y$ and $\Phi$ yields $D_s\Pi=0$ and $D_n\Pi=0$.} to satisfy
\begin{eqnarray}
    D_s\Pi =0 \, , \qquad D_n\Pi=0 \, , \qquad \Pi \equiv \frac{rf H_2}{2q^3} = \frac{\delta m}{q^3} \, .
\end{eqnarray}
In fact, $\Pi$ is the conjugate momentum to $\Phi$ and $\delta m$ is the perturbation of the mass function~\eqref{massfunction}, defined by $g^{\mu\nu}\partial_\mu r\,\partial_\nu r = 1 - 2m/r$. Hence $\Pi$ is constant and there is no local propagating scalar mode on this branch. This conclusion relies on $X'\neq0$ and does not apply to constant $X$ stealth solutions of k-essence, for which $P_{XX}$ need not vanish~\cite{Mukohyama:2005rw,Mukohyama:2025owu}.

\medskip

We now turn to the nontrivial case $P_{\cal F}\neq0$. For ${\cal F}<0$, requiring $\mathcal A$ to vanish identically gives the family~\eqref{eq:exceptional-P}
\begin{equation}
	P(X,{\cal F})=K(X)+Q(X)\sqrt{-{\cal F}}\,.
\end{equation}
The associated background equations were derived in the subsection~\ref{sec:exceptional-background}.

 For monopole perturbations, we follow the same strategy as in the generic case and we return to~\eqref{eq:quadratic-action-monopole-Phi} and set $\mathcal A=0$. Integrating by parts, we find that $\Phi$ and $\mathcal Y$ appear without derivatives and can both be integrated out. We are left with only one variable\footnote{It is interesting to notice that although $\delta\phi=0$ in the gauge we are using here, the scalar invariant $X$ can still fluctuate through the metric perturbations. Indeed, the $\Phi$ equation of~\eqref{eq:quadratic-action-monopole-Phi} gives
\begin{equation}
	\delta X=-\mathcal Y
	=\frac{\sqrt{-X}}{2\sigma_X q r Q f\psi'}D_nU\,.
	\label{eq:exceptional-deltaX}
\end{equation}
Thus the remaining scalar degree of freedom is characterized by the metric variable $U$ in this gauge.} $U \equiv rfH_2$ whose dynamics is governed by the  quadratic action 
\begin{equation}
	S_{\mathrm{quad}}[U]=\int\D \tau\,\D r\,\D \Omega\,
	\mathcal K\left[(D_nU)^2+v_s(D_nU)(D_sU)\right]\,,
	\label{eq:exceptional-U-action}
\end{equation}
where $D_n$ and $D_s$ are defined in~\eqref{eq:exceptional-frame}, and
\begin{equation}
	\mathcal K=\frac{(rX)'}{4qr^2(f\psi')^2|X'|Q}\,,\qquad
	v_s=\frac{rf\psi'X'}{q(rX)'}\,.
	\label{eq:exceptional-coefficients}
\end{equation}

To study the stability of the monopole perturbations, we compute the Hamiltonian. As shown in Appendix~\ref{sec:monopole-hamiltonian-exceptional}, it is convenient to express the Lagrangian density in $(\tau,r)$ coordinates
\begin{equation}
 \mathcal L=-\frac{X}{q^2}\,\mathcal K(\partial_\tau U+u_0\partial_rU)(\partial_\tau U+u_1\partial_rU)\,,
 \label{LagdensityU}
\end{equation}
from which we can easily compute the canonical momentum $\Pi_U$ and express the Hamiltonian density given by \eqref{eq:exceptional-H-density}
\begin{equation}
 \begin{aligned}
 \mathcal H
 &= \frac{\big[q^2\Pi_U+X\,\mathcal K(u_0+u_1)\partial_rU\big]^2}{-4q^2X\,\mathcal K}
 +\frac{(\partial_rU)^2}{4qr^2|X'|Q} \,.
 \end{aligned}
\end{equation}
The two coordinate velocities $u_0$ and $u_1$ are defined by 
\begin{equation}
 u_0\equiv\frac{qf\psi'}{X}\,,\qquad
 u_1\equiv\frac{qf\psi'}{(rX)'}\,,
 \label{eq:exceptional-H-definitions-body}
\end{equation}
and their physical interpretation will become clear in the following subsection.
As a consequence, for $q>0$,  this Hamiltonian is non-negative and thus bounded from below if and only if
\begin{equation}
\mathcal K>0\,,\qquad Q>0
 \qquad\Longleftrightarrow\qquad
 Q>0\,,\qquad (rX)'>0\,.
 \label{eq:exceptional-H-positivity}
\end{equation}
These conditions provide a sufficient energy criterion for linear monopole stability and they are satisfied in particular for the black hole solutions~\eqref{eq:BH-special-branch} under specific constraints to the $\alpha$ and $\beta$ parameters.
Let us remark that positivity of the Killing energy is not necessary for bounded classical evolution~\cite{Deffayet:2021nnt}. 

\subsubsection{Dynamics of the monopole }

To gain further insight into the dynamics of the monopole, it is useful to interpret its propagation in terms of a two-dimensional effective geometry. Indeed, the Lagrangian density \eqref{LagdensityU} can be recast as
\begin{eqnarray}
    \mathcal L = \mathcal Z(r)\sqrt{\left|g^{\rm (m)}\right|}\,
    g^{\rm (m)\mu\nu}\partial_\mu U\,\partial_\nu U\,,
\end{eqnarray}
where \(\mathcal Z(r)\) is a position-dependent coupling. Its presence is essential since, in two dimensions, the kinetic term of a minimally coupled scalar field is conformally invariant. The effective metric is therefore defined up to a conformal factor, while the radial dependence of the overall kinetic coupling cannot be absorbed into it.

A convenient representative of the effective metric, in the \((\tau,r)\) coordinates, is
\begin{eqnarray}
    \D s_{\rm (m)}^2
    =g^{\rm (m)}_{\mu\nu}\D x^\mu \D x^\nu
    = \frac{2}{(u_0-u_1)^2}
    \left(\D r-u_0\D \tau\right)
    \left(\D r-u_1\D \tau\right)\,,
\end{eqnarray}
whose components are
\begin{eqnarray}
    g^{\rm (m)}_{\tau\tau}=\frac{2u_0u_1}{(u_0-u_1)^2}\,,\qquad
    g^{\rm (m)}_{\tau r}=-\frac{u_0+u_1}{(u_0-u_1)^2}\,,\qquad
    g^{\rm (m)}_{rr}=\frac{2}{(u_0-u_1)^2}\,.
    \label{matrixeffelements}
\end{eqnarray}
The corresponding kinetic coupling is
\begin{equation*}
\mathcal Z(r)=\frac{X\mathcal K}{2q^2}|u_0-u_1|\,,
\end{equation*}
while the determinant of the effective metric reads
\begin{eqnarray}
    g^{\rm (m)}
    =-\frac{1}{(u_0-u_1)^2}\,,
\end{eqnarray}
which shows that the metric has Lorentzian signature whenever \(u_0\neq u_1\).

The two null directions of this geometry are immediately identified as
\begin{eqnarray}
   \D r=u_0\,\D\tau\,\qquad \text{and} \qquad
   \D r=u_1\,\D\tau\,.
\end{eqnarray}
Thus, \(u_0\) and \(u_1\) directly determine the two characteristic directions along which monopole perturbations propagate. Hence, from~\eqref{eq:exceptional-frame}, the direction $\D r=u_0\,\D\tau$ follows the background scalar field flow. Interestingly, when the positive energy conditions discussed above~\eqref{eq:exceptional-H-positivity} are satisfied, one has \(u_0u_1<0\). The two characteristic directions then correspond to propagation towards opposite radial directions.

Another interesting feature concerns the stationary Killing vector \(\partial_\tau\) whose norm in the effective geometry is $g^{\rm (m)}_{\tau\tau}$ \eqref{matrixeffelements}. Hence,
in the positive energy branch, where \(X<0\), \(\mathcal K>0\), and \(u_0u_1<0\), this quantity is negative. Hence, the Killing vector remains timelike with respect to the effective metric. This suggests that the monopole dynamics does not exhibit an ergoregion in this branch.

However, the situation may change if \(u_0u_1\) changes sign. In particular, using
\begin{eqnarray}
    u_0u_1=\frac{q^2f^2\psi'^2}
    {X(rX)'}\,,
\end{eqnarray}
we see that, since \(X<0\), a region where \((rX)'<0\) corresponds to \(u_0u_1>0\). In such a region, \(\partial_\tau^2>0\)  and the stationary Killing vector becomes spacelike with respect to the effective metric. 
{This raises the question of whether the negative Killing energy on the
branch with $Q<0$ and $(rX)'<0$ admits an ergoregion interpretation.}  We hope to study this point in the near future. 

\section{Summary and conclusions}
\label{sec:Summary}

We studied static, spherically symmetric solutions in F-UDHOST gravity with scalar sector $P(X,{\cal F})$ and a scalar field linear in time $\phi = q\, t + \psi(r)$. On the generic branch $q\psi'\neq0$, the background equations admit a simple point particle description in terms of the scalar kinetic invariant $X(r)$. Once $X(r)$ is determined, the metric follows by direct integration. This gives a systematic way to construct exact backgrounds and to see how the scalar interactions modify the Schwarzschild geometry. The examples considered here include Yukawa-like, inverse-square and more general power-law corrections.

We find that, given the shape of the action, axial perturbations behave similarly in F-UDHOST as in GR, and that they are stable. However, we find that the monopole sector of even perturbations has a pathology in general. Indeed, considering nonextremal black holes on the regular generic branch at short wavelengths, we find that the quadratic Killing Hamiltonian is unbounded while the sound speed vanishes at every finite radius outside the horizon independently of the asymptotic geometry. This result assumes $q\psi'X'\neq0$, a timelike scalar gradient $X<0$, and a regular horizon. While we observe that the choice of variable can change the relative signs in the quadratic action, the quadratic Killing Hamiltonian remains unbounded from below in every case. Finally, the vanishing sound speed also raises a strong coupling concern~\cite{Motohashi:2019ymr,Takahashi:2021bml}. The \textit{scordatura} mechanism can restore spatial propagation~\cite{DeFelice:2022xvq,Mukohyama:2025owu}. Whether the quadratic monopole  Killing energy is then bounded from below should be checked on a black hole solution of the modified theory.

The special family of theories $P(X,{\cal F})=K(X)+Q(X)\sqrt{-{\cal F}}$ can evade this instability because $\mathcal A=0$ identically for ${\cal F}<0$. We derive the general background construction and exact quadratic monopole action for this family, and obtain the conditions for positive Killing energy and real radial propagation speeds in this sector. As an illustration, we construct an explicit family of black holes (together with the scalar field) that satisfy these monopole conditions. Studying their asymptotic behavior and stability beyond the monopole sector are interesting directions for future work.

\subsubsection*{Acknowledgements}
The work of MAG and PP was supported by IBS under the project code, IBS-R018-D3. The work of KN is partially supported by ANR grant StronG (No. ANR-22-CE31-0015-01). The work of HR is supported by the Alexander von Humboldt foundation.

\noindent\textbf{Statement on the use of AI.}
The authors used ChatGPT (OpenAI, Astra 6.2) and Claude (Anthropic, Opus 5.5 and Fable 5.2) to assist with symbolic calculations. They independently verified the results and take full responsibility for the accuracy and integrity of the article.

\appendix

\section{On the field equations}
\label{app:field-equations}
In this appendix, we give some details concerning the equations of motion for the background in their covariant forms and for static and spherically symmetric configurations. 

\subsection{Covariant form}

We start by considering a general action of the form
    \begin{align}\label{eq:actiongeneral}
	S[g_{\mu\nu},\phi] & \equiv \int \mathrm{d}^4 x \, \sqrt{-g}\,
	\left[
	\frac{M_{\rm Pl}^2}{2}\, R
	+ {L}\!\left(X,Y,Z\right)
	\right] \, ,
\end{align}
where $L$ is an arbitrary function of the variables $X$, $Y$ and $Z$ at this stage. We first focus on the equations of the metric tensor. After a direct calculation, we show that
\begin{eqnarray}
   {\cal E}_{\mu\nu} =  \Mpl^2 \, G_{\mu\nu} - T_{\mu\nu} \, = 0 \, ,
\end{eqnarray}
where the stress-energy tensor reads
\begin{eqnarray}
\label{eq:stress-tensor-general}
    T_{\mu\nu} = L g_{\mu\nu}  -2 \left[
    \mathcal Q \, \phi_\mu \phi_\nu + L_Y \, \phi_{(\mu} X_{\nu)} + L_Z \, X_\mu X_\nu 
    \right] \, , 
\end{eqnarray}
where we have introduced
\begin{eqnarray}
\mathcal Q=L_X + \nabla_\alpha H^\alpha \, , \qquad   H_\alpha = - L_Y \, \phi_\alpha - 2 L_Z \, X_\alpha \, ,
\end{eqnarray}
and we use the standard notation $L_X$ for the derivative of $L$ with respect to $X$ and similarly for $Y$ and $Z$. Notice that we are using the convention $A_{(\mu}B_{\nu)} = (A_\mu B_\nu + A_\nu B_\mu)/2$ for the symmetrization between any two tensors $A_\mu$ and $B_\mu$.

The equation for the scalar field, which is can be deduced from these equations due to the invariance under diffeomorphisms, takes the expected form
\begin{align}
\nabla_\alpha J^{\alpha} = 0 \, ,
\end{align}
where $J_\mu$ is the Noether current associated with the shift symmetry. It is explicitly given by
\begin{eqnarray}
    J_\mu \; = \; \mathcal Q \, \phi_\mu + \frac{1}{2} L_Y \, X_\mu  \, .
\end{eqnarray}
In the case we are considering in the paper, where $L=P(X,{\cal F})$ with ${\cal F}=XZ-Y^2$, we have
\begin{eqnarray}
    L_X = P_X + Z P_{\cal F} \, , \qquad
    L_Y = -2Y P_{\cal F} \, , \qquad
    L_Z = X P_{\cal F} \, ,
\end{eqnarray}
so that the stress-energy tensor becomes
\begin{eqnarray}
\label{eq:stress-tensor-P}
    T_{\mu\nu} = P g_{\mu\nu} - 2 \left[
    \mathcal Q \,  \phi_\mu \phi_\nu + (X  \, X_\mu X_\nu -2 Y  \, \phi_{(\mu} X_{\nu)} )P_{\cal F} 
    \right] 
\end{eqnarray}
	where  $\mathcal Q$ and $H_\alpha$ take now the form
    \begin{eqnarray}
        \mathcal Q= \, P_X + Z P_{\cal F} + \nabla_\alpha H^\alpha \, , \qquad  \quad  H_\alpha = 2P_{\cal F} (Y \phi_\alpha - X X_\alpha) \, .
    \end{eqnarray}
For $X<0$, the general energy-momentum tensor~\eqref{eq:stress-tensor-general} admits an imperfect fluid decomposition~\cite{Gorji:2025ajb}. For the subset~\eqref{eq:stress-tensor-P}, the energy flux vanishes in the scalar frame.

\subsection{Symmetry reduction}
In this subsection, we assume the geometry to be static and spherically symmetric, while the scalar field admits a linear time dependency \eqref{eq:spherical background}.  
In that case, the expressions of $X$, $Y$ and $Z$ simplify,
\begin{eqnarray}
    X=-\frac{q^2}{f} + h \psi'^2 \, , \quad 
    Y= h \psi' X' \, , \quad 
    Z= h X'^2 \, .
\end{eqnarray}
and $\mathcal Q$ becomes
\begin{eqnarray}
    \mathcal Q = L_X - \frac{1}{r^2} \sqrt{\frac{h}{f}} \frac{\D}{\D r} \left[r^2 \sqrt{fh} X' \left( \frac{Y L_Y}{Z} + 2 L_Z \right)  \right] \, .
\end{eqnarray}

The above Einstein equations reduce to 4 independent differential equations for the three unknowns $f$, $h$ and $\psi$. To lighten the notations, we consider units where $\Mpl=1$ from now on.  The first and the simplest equation is given by the non-trivial non-diagonal Einstein equation which reduces to
\begin{eqnarray}
    T_{rt} = -q \left(2 \psi' \mathcal Q + X' L_Y\right) = 0 \, .
\end{eqnarray}
In the generic case where $q\neq 0$, $\psi' \neq 0$ and $X' \neq 0$, which we assume here, this equation can be equivalently reformulated as follows
\begin{eqnarray}
    2 \mathcal Q Y + Z L_Y = 0 \, .\label{Ert}
\end{eqnarray}
The 3 diagonal Einstein equations are given by
\begin{eqnarray}
    {\cal E}_t^t  & = & \frac{h-1}{r^2} + \frac{h'}{r} - L + \frac{q^2 }{f } \frac{Z L_Y}{Y} \; = \; 0   \, ,\label{Ett} \\
    {\cal E}_r^r  & = & \frac{h-1}{r^2} + \frac{h}{r} \frac{f'}{f} - L +  Y L_Y  + 2 Z L_Z   \; = \; 0 \, ,\label{Err}  \\
    {\cal E}_\theta^\theta & = & {\cal E}_\varphi^\varphi = \frac{h f''}{2 f} - \frac{hf'^2}{4 f^2} + \frac{h'f'}{4f} + \frac{hf'}{2rf} + \frac{h'}{2r} - L \; = \, 0 \, . \label{Etheta} 
\end{eqnarray}
Bianchi identities insures that these 3 equations are non-independent and we can express the last one \eqref{Etheta} in terms of \eqref{Ert}, \eqref{Ett} and \eqref{Err}. Hence we are left with 3 equations for 3 unknowns.

To solve the system, we start with the following combination 
\begin{eqnarray}
    {\cal E}_t^t - {\cal E}_r^r = \frac{h}{r} \left( \frac{h'}{h} - \frac{f'}{f} \right) - \frac{Z}{Y}(2Y L_Z + X L_Y) \, = 0 \, .
\end{eqnarray}
As a consequence, when $X' \neq 0$, the solution is homogeneous, i.e. $f=\alpha h$ (where $\alpha$ is an irrelevant  constant that can be fixed to 1 without loss of generality), 
{whenever the following condition holds identically:}
\begin{eqnarray}
    2YL_Z+ XL_Y = 0 \; \Longleftrightarrow \; L(X,Y,Z)=P(X,XZ-Y^2) \, ,
\end{eqnarray}
where $P$ is an arbitrary function. Hence, interestingly, the form of the action \eqref{eq:action} is directly linked to the fact that the background metric is homogeneous. To get a non-homogeneous solution, one must consider a more general Lagrangian \eqref{eq:actiongeneral}. 

As a consequence, when $L=P(X,{\cal F})$, we have $f=h$ and the equation \eqref{Ert} simplifies into
\begin{eqnarray}
\label{eqYX}
    Y\left[r^2 P_X - (r^2 P_{X'})'\right] = 0 \, , 
\end{eqnarray}
which reproduces the expected equation \eqref{eq:SF-master}  for the scalar $X(r)$ when $X'\neq 0$.

\section{On perturbations}
Here we provide some details concerning the dynamics of perturbations. We use $\Mpl=1$ as in Sec.~\ref{sec:perturbations}.
\subsection{Axial perturbations}
\label{App:perturbations}
Now we provide further details on the dynamics of axial perturbations. Following the conventions \eqref{eq:odd-perttext} and after a Fourier transform \eqref{Fourtrans}, the two equations $\delta {\cal E}_\theta^\theta$ and $\delta {\cal E}_r^\theta$ for the perturbations in a background, which is not necessarily homogeneous, are given by
\begin{eqnarray}
    \omega h_0  - i \sqrt{fh} \left( \sqrt{fh} h_1\right)' = 0 \, , \label{firstpert} 
\end{eqnarray} 
and 
\begin{eqnarray}
    h_0' - \frac{2 h_0}{r} + i\omega h_1 - \frac{i f}{\omega r^2} \left( 2 \lambda_\ell + \frac{r(fh)'}{f} + \frac{r^2 h f''}{f} - 2r^2 L \right)h_1 
    + \frac{f' \left( f h' - f' h\right)}{2i\omega f} h_1
    \, = 0 \, , 
\end{eqnarray}
where $L$ is evaluated in the background. If we replace $L$ by its expression in terms of the components of the metric using \eqref{Etheta}, the second equation drastically simplifies and leads to
\begin{eqnarray}
    h_0' - \frac{2 h_0}{r} + i\omega h_1 - \frac{2i\lambda_\ell f}{r^2} \frac{h_1}{\omega} \; = \; 0 \, . \label{secondpert}
\end{eqnarray}
Finally, we can combine the two equations \eqref{firstpert} and \eqref{secondpert} to form the  system of first order differential equations
\begin{eqnarray}
\frac{\D}{\D r}
    \begin{pmatrix}
        h_0 \\
        h_1/\omega
    \end{pmatrix} = 
    \begin{pmatrix}
        2/r & -i \omega^2 + 2i\lambda_\ell f/r^2 \\
        -i/(fh) & -(f'/f+h'/h)/2
    \end{pmatrix}
     \begin{pmatrix}
        h_0 \\
        h_1/\omega
    \end{pmatrix} \, ,
\end{eqnarray}
which reproduces exactly the dynamics of a massless spin 2 field in the background metric, as shown in \cite{Langlois:2022ulw}. 

As expected, the effective metric of the axial perturbations coincides with the background metric since the theory is defined in an Einstein frame without non-minimal coupling to the Ricci scalar in the action. This result still hold when $L$ is any Lagrangian not necessarily in the form $P(X,{\cal F})$.

\subsection{Coefficients entering the monopole lagrangians}
\label{app:Quad_Action_Monopole}
We give here the values of the coefficients appearing in the various Lagrangians of Sec.~\ref{sec:Monopole}, assuming $X' \neq 0$. In first place we define the $c_i$ appearing in~\eqref{eq:Monopole_quad_action}:
\begin{subequations}
\begin{alignat}{2}
    c_1 &= f X' \psi'\,,
    &\qquad c_2 &= -\frac{(2q^2+fX)X'}{2q}\,,\\[4pt]
    c_3 &= -X'\big(4q^2P_{\cal F}+r\mathcal{A}\zeta\big)\,,
    &\qquad c_4 &= 2(rf)'+\frac{f^2}{q^2}(X-rX')\,,\\[4pt]
    c_5 &= 4P_{\cal F}(q^2-rfX')-\frac{r^2\mathcal{A}\zeta^2}{2q^2}
           -(r^2\mathcal{A}\zeta)'\,,
    &\qquad \zeta &= fX'+\frac{2q^2X''}{X'}\,.
\end{alignat}
\end{subequations}
We further define the coefficients appearing in~\eqref{eq:quadratic-action-monopole-Phi} and~\eqref{eq:quadratic-action-monopole-reduced}:
\begin{subequations}
    \begin{align}
        d_0 &= X^2\Big(\frac{c_4}{4}-r^2\mathcal A c_2^2\Big) \,,\\
        d_1 &= -\frac{1}{\sqrt{-X}}
          \Big[X\Big(\frac{f\psi'}X\Big)'
          +r\mathcal A\psi'(2q^2+fX)X'^2\Big] \,,\\
        d_2 &= \sqrt{-X}\Big[\frac{rfX'}{2X^2}
          -\frac{r(2q^2+fX)}{2q^2}
           \Big(c_3+\frac{r\mathcal A(q^2+fX)X'^2}{X}\Big)\Big] \,,\\
        d_3 &= \frac r{2q}
          \Big[f+2r\mathcal AX(2q^2+fX)X'\Big] \,,\\
        d_4 &= \frac{(q^2+fX)X'}{2qX\sqrt{-X}} \,,\\
        d_5 &= \frac{f\psi'X'}{2qX\sqrt{-X}} \,,\\
        d_6 &= -\frac{rf\psi'c_3}{2q^2X^2} \,,\\
        d_7 &= \frac{1}{2(-X)}
          \Big[c_5-\frac{r(q^2+fX)}{q^2X}c_3\Big] \,.
    \end{align}
\end{subequations}
Here $d_0/X^2$ is the total coefficient of $H_2^2$ in~\eqref{eq:quadratic-action-monopole-Phi}, including the contribution of $\mathcal W^2$.

\section{Hamiltonian structure of the monopole sector}
\label{sec:monopole-hamiltonian}

In this section we revisit the monopole sector from the Hamiltonian point of view. This makes the counting of degrees of freedom explicit, identifies the canonical variables, and provides an independent derivation of the results of subsections~\ref{sec:Monopole} and~\ref{sec:exceptional}. 

\subsection{Non-reduced phase space and canonical Hamiltonian}

We use the coordinates $(\tau,r)$ and take the hypersurfaces $\tau=\text{const}$, i.e.\ the constant-$\phi$ hypersurfaces, as Cauchy surfaces. They are spacelike since $X<0$.
Moreover, $\partial_\tau|_r=\partial_t|_r$ is the background Killing vector, so that the Hamiltonian constructed below generates Killing time translations.

The only time derivative in the action~\eqref{eq:quadratic-action-monopole-Phi} is $\partial_\tau\Phi$, which appears linearly and multiplied by $H_2$. After an integration by parts, the action is therefore already in first-order form,
\begin{equation}
 S_\text{quad}=\int\D\tau\,\D r\,\D\Omega\,\big[\Pi\,\partial_\tau\Phi-\mathcal H\big]\,,
 \qquad
 \Pi\equiv \frac{rf}{2q^3}H_2=\frac{\delta m}{q^3}\,,
 \label{eq:monopole-first-order}
\end{equation}
with the Hamiltonian density
\begin{equation}
\begin{aligned}
 \mathcal H={}&r^2\mathcal A\,\mathcal W^2
 +\frac qX\big(f\psi'\Phi-q^2\mathcal Y\big)\partial_r\Pi
 -\frac{q^6c_4}{r^2f^2}\Pi^2\\
 +&\frac{q\,c_3(2q^2+fX)}{f}\,\Pi\,\mathcal Y
 -\frac{r\,c_3f\psi'}{2q^2X}\,\Phi\,\mathcal Y
 +Xd_7\,\mathcal Y^2\,,
\end{aligned}
\label{eq:monopole-hamiltonian}
\end{equation}
where $\mathcal W$ is now expressed in terms of $\Pi$,
\begin{equation}
 \mathcal W=-q\partial_r\mathcal Y-\gamma\,\mathcal Y+\alpha\,\Phi+\frac{2q^3c_2}{rf}\,\Pi\,,
 \qquad
 \alpha\equiv\frac{f\psi'X'}{2qX}\,,\qquad
 \gamma\equiv\alpha f\psi'=\frac{(fX+q^2)X'}{2qX}\,.
 \label{eq:monopole-W}
\end{equation}
In~\eqref{eq:monopole-first-order}, $\delta m$ is the perturbation of the mass function~\eqref{massfunction}, defined by $g^{\mu\nu}\partial_\mu r\,\partial_\nu r = 1 - 2m/r$.

\subsection{Constraints analysis}

The variables $(\mathcal Y,\Phi,H_2)$ come with their conjugate momenta $(\pi_{\mathcal Y},\pi_\Phi,\pi_{H_2})$, and the primary constraints are
\begin{equation}
 \chi_1\equiv\pi_\Phi-\Pi\approx0\,,\qquad
 \chi_2\equiv\pi_{H_2}\approx0\,,\qquad
 \chi_3\equiv\pi_{\mathcal Y}\approx0\,,
\end{equation}
where $\approx$ holds for weak equality (i.e. equality up to constraints). 

Now we have to study the stability of these constraints under time evolution to see whether they generate new constraints or not. For that, we first introduce the total Hamiltonian density
\begin{eqnarray}
    \mathcal{H}_{T} = \mathcal{H} + \lambda_1 \chi_1+ \lambda_2 \chi_2 + \lambda_3 \chi_3 \, ,
\end{eqnarray}
where $\lambda_i$ are Lagrange multipliers imposing the primary constraints. Since 
\begin{eqnarray}
    \{\chi_1(r),\chi_2(r')\}=-\tfrac{rf}{2q^3}\,\delta(r-r') \, ,
\end{eqnarray}
the stability of $\chi_1$ and $\chi_2$ leads to fixing the Lagrange multipliers $\lambda_1$ and $\lambda_2$, hence it does not create new constraints. Furthermore, $(\chi_1,\chi_2)$ forms a subset of 
second class constraints. 
We can solve the two constraints explicitly by eliminating $\pi_{H_2}$ and expressing $H_2$ by $H_2=2q^3\Pi/(rf)$. As a consequence, the total Hamiltonian density reduces to 
\begin{eqnarray}
   \mathcal H+\lambda_3\,\pi_{\mathcal Y}
   \end{eqnarray}
 Stability of $\pi_{\mathcal Y}$ leads to the secondary constraint
\begin{equation}
\begin{aligned}
 \mathcal C\equiv
{}&2q\,\partial_r\big(r^2\mathcal A\,\mathcal W\big)
 -2r^2\mathcal A\,\gamma\,\mathcal W
 -\frac{q^3}{X}\,\partial_r\Pi\\
 +&\frac{q\,c_3(2q^2+fX)}{f}\,\Pi
 -\frac{r\,c_3f\psi'}{2q^2X}\,\Phi
 +2Xd_7\,\mathcal Y\approx0\,.
\end{aligned}
\label{eq:monopole-secondary}
\end{equation}
Its bracket with the primary constraint can be shown to be non-trivial. As a consequence, the stability of $\mathcal{C}$ leads to a differential equation for the remaining Lagrange multiplier $\lambda_3$ which fixes it (with appropriate boundary conditions).

As a consequence, the Dirac analysis stops here with 4 second class constraints for an initial 6-dimensional non-reduced phase space. Hence, we end up with the conclusion that there is only one physical degree of freedom which corresponds to the expected monopole. 

Few remarks are in order. First, the constraint~\eqref{eq:monopole-secondary}  is an ordinary differential equation in $r$  which determines $\mathcal Y$ in terms of $(\Phi,\Pi)$ and of the boundary conditions. This is the Hamiltonian counterpart of the statement that $\mathcal Y$ is a shadowy mode~\cite{DeFelice:2018ewo}. Second, the assumption $d_0\neq0$ needed to obtain~\eqref{eq:quadratic-action-monopole-reduced} is therefore an artefact of the Lagrangian reduction, which explains why $d_0$ drops out of~\eqref{eq:high-k-action}. 

\subsection{Generic branch at short wavelengths}
\label{sec:monopole-hamiltonian-generic}

We now take the high wavenumber $k$ limit of subsection~\ref{sec:Monopole}, with $\Phi,\mathcal Y$ of order 1 while radial derivatives are of order $k$. Hamilton's equation for $\Phi$ then shows that $\Pi$ is of order $k$. At leading order, the constraint~\eqref{eq:monopole-secondary} leads to
\begin{equation}
 -q\,\partial_r\mathcal Y+\frac{2q^3c_2}{rf}\,\Pi\simeq\frac{q^2}{2r^2\mathcal AX}\,\Pi\,.
 \label{eq:monopole-Y-solution}
\end{equation}
Substituting this solution into~\eqref{eq:monopole-hamiltonian} gives, at order $k^2$, the reduced Hamiltonian density\footnote{Performing the Legendre transform back, we obtain $\mathcal L\simeq q^2(D_n\Phi)^2/[4(-X)\mathcal N]$, which coincides with~\eqref{eq:high-k-action} by virtue of~\eqref{eq:monopole-N}. }
\begin{equation}
 \mathcal H_\text{red}\simeq\mathcal N\,\Pi^2-u\,\Pi\,\partial_r\Phi\,,
 \qquad u\equiv\frac{qf\psi'}{X}\,,
 \label{eq:monopole-Hred}
\end{equation}
with
\begin{equation}
 \mathcal N=-\frac{q^4}{4r^2\mathcal AX^2}+\frac{2q^5c_2}{rfX}-\frac{q^6c_4}{r^2f^2}
 =-\frac{q^4\,\Delta}{4r^2f^2\mathcal AX^2}\,.
 \label{eq:monopole-N}
\end{equation}
The second equality follows from the expressions of $c_2$ and $c_4$ given in Appendix~\ref{app:Quad_Action_Monopole}, which satisfy the algebraic identity
\begin{equation}
 q^2X\,c_4-2qrf\,c_2=2q^2(rfX)'+f^2X^2\,.
 \label{eq:c2c4-identity}
\end{equation}
Equivalently, $\Delta=f^2+4\mathcal AX\,(q^2Xc_4-2qrf\,c_2)$. 

{Under the assumptions of Sec.~4.2.3, Eqs.~\eqref{eq:monopole-horizon-ratio}
and~\eqref{eq:monopole-N} give $\mathcal N<0$ throughout the connected
static exterior for our choice of variables $(\Phi,\Pi)$.
On this branch, $u=qf\psi'/X\ne0$. The determinant of the quadratic form
in $(\Pi,\partial_r\Phi)$ in Eq.~\eqref{eq:monopole-Hred} is therefore
$-u^2/4<0$. Thus the short-wavelength quadratic Killing Hamiltonian is
unbounded from below for either sign of $\mathcal N$.}

One last remark concerning the equations of motion. Since $\partial_\tau+u\,\partial_r=(q/\sqrt{-X})\,D_n$, Hamilton's equations take the form
\begin{equation}
 D_n\Pi\simeq0\,,\qquad
 D_n\Phi\simeq\frac{2\sqrt{-X}}{q}\,\mathcal N\,\Pi\,,
 \label{eq:monopole-Hamilton-generic}
\end{equation}
so that $D_n^2\Phi\simeq0$. The principal symbol has a double root, and the corresponding characteristic is the integral curve of $n^\mu$. This is the precise sense in which the sound speed vanishes.

\subsection{Special branch}
\label{sec:monopole-hamiltonian-exceptional}

On the special branch $\mathcal A=0$, the constraint structure simplifies as we are going to show. 

We first notice that, for the family~\eqref{eq:exceptional-P}, the background relation $P_{\cal F}=-\sigma_XQ/(2qX')$ in~\eqref{eq:exceptional-bg}, together with the coefficients of Appendix~\ref{app:Quad_Action_Monopole}, yields
\begin{equation}
 c_3=2\sigma_X q\,Q\,,\qquad
 d_7=\frac{q\,Q\,(rX)'}{X^2|X'|}\,.
 \label{eq:exceptional-c3-d7}
\end{equation}
Furthermore, when $\mathcal A=0$, the term $r^2\mathcal A\,\mathcal W^2$ obviously vanishes from~\eqref{eq:monopole-hamiltonian}. After radial integration by parts, $\Phi$ and $\mathcal Y$ enter the Hamiltonian algebraically, while $\Pi$ appears through its radial derivative. In particular, the secondary constraint~\eqref{eq:monopole-secondary} becomes algebraic in $\mathcal Y$. Provided $d_7\neq0$, or equivalently $Q(rX)'\neq0$ on the regular branch, the pair $(\pi_{\mathcal Y},\mathcal C)$ remains second class. The counting of degrees of freedom is therefore unchanged, leaving a single propagating mode.

The difference with the generic branch lies in the choice of dynamical variable. On the generic branch, $\mathcal H$ is quadratic in $\Pi$, which can be eliminated, leaving $\Phi$ as the coordinate. On the special branch, $\Phi$ appears only linearly in~\eqref{eq:monopole-hamiltonian} and acts as a Lagrange multiplier, while $\Pi$ appears with radial derivatives. It is therefore $\Pi$ that should be kept\footnote{This amounts to the canonical transformation $(\Phi,\Pi)\to(\Pi,-\Phi)$. It preserves the symplectic form, since $\Pi\,\partial_\tau\Phi=-\Phi\,\partial_\tau\Pi$ up to a total derivative.}. Since $\Pi=U/(2q^3)$ with $U=rfH_2=2\,\delta m$, the propagating degree of freedom on this branch is described by the mass perturbation. 

After some calculations\footnote{Using $c_3=2\sigma_XqQ\neq0$, we easily show that the Hamilton equation for $\Pi$, determines $\mathcal Y$:
\begin{equation}
 \mathcal Y=\frac{2q^2X}{r\,c_3f\psi'}\,\big(\partial_\tau+u\,\partial_r\big)\Pi
 =-\frac{\sqrt{-X}}{2\sigma_Xq\,r\,Q\,f\psi'}\,D_nU\,,
 \label{eq:exceptional-Y-Hamilton}
\end{equation}
in agreement with~\eqref{eq:exceptional-deltaX}, while the secondary constraint determines $\Phi$. Substituting these relations into the canonical action yields
\begin{equation}
 \begin{aligned}
 \mathcal L={}&\frac{q^3}{X}\,\mathcal Y\,\partial_r\Pi
 -\frac{q\,c_3(2q^2+fX)}{f}\,\Pi\,\mathcal Y -X d_7\,\mathcal Y^2
 +\frac{q^6c_4}{r^2f^2}\,\Pi^2\,,
 \end{aligned}
\end{equation}
where $\mathcal Y$ is given by~\eqref{eq:exceptional-Y-Hamilton}.

Finally, the identity
\begin{equation}
 \frac{q^2c_4}{r^2f^2}=-\left(\frac{2q^2+fX}{rf}\right)'.
 \label{eq:c4-identity}
\end{equation}
enables us to recover  the  reduced action~\eqref{eq:exceptional-U-action}.}, we show that solving the constraints leads to the exact reduced action~\eqref{eq:exceptional-U-action}, with
\begin{equation}
 \mathcal K=\frac{X^2d_7}{r^2c_3^2(f\psi')^2}\,,\qquad
 v_s=\frac{r\,c_3f\psi'}{2qX^2d_7}\,.
\end{equation}
Using~\eqref{eq:exceptional-c3-d7}, these expressions coincide with~\eqref{eq:exceptional-coefficients}. In particular, the no-ghost condition $\mathcal K>0$ is equivalent to $d_7>0$.

\medskip

We now turn to the Hamiltonian associated with this reduced action. Unlike in the generic branch, the Legendre transformation can be performed exactly. Introducing
\begin{equation}
 u_0\equiv\frac{qf\psi'}{X}=u\,,\qquad
 u_1\equiv\frac{qf\psi'}{(rX)'}\,,
 \label{eq:exceptional-H-definitions}
\end{equation}
the frame derivatives in~\eqref{eq:exceptional-frame} satisfy
\begin{equation}
 \begin{aligned}
 D_nU&=\frac{\sqrt{-X}}{q}(\partial_\tau U+u_0\partial_rU)\,,\qquad
 D_nU+v_sD_sU=\frac{\sqrt{-X}}{q}(\partial_\tau U+u_1\partial_rU)\,.
 \end{aligned}
\end{equation}
The reduced Lagrangian therefore takes the factorized form
\begin{equation}
 \mathcal L=-\frac{X}{q^2}\,\mathcal K(\partial_\tau U+u_0\partial_rU)(\partial_\tau U+u_1\partial_rU)\,.
 \label{eq:exceptional-H-Lagrangian}
\end{equation}
The corresponding canonical momentum is
\begin{equation}
 \Pi_U=\frac{-X}{q^2}\,\mathcal K\left[2\partial_\tau U+(u_0+u_1)\partial_rU\right]\,.
 \label{eq:exceptional-H-momentum}
\end{equation}
Performing the Legendre transformation, we obtain the exact quadratic Hamiltonian density generating translations along the background Killing vector $\partial_\tau$:
\begin{equation}
 \begin{aligned}
 \mathcal H
 &=\Pi_U\partial_\tau U-\mathcal L\\
 &=\frac{\big[q^2\Pi_U+X\,\mathcal K(u_0+u_1)\partial_rU\big]^2}{-4q^2X\,\mathcal K}
 +\frac{(\partial_rU)^2}{4qr^2|X'|Q}\,.
 \end{aligned}
 \label{eq:exceptional-H-density}
\end{equation}
In the last equality, we used $X\,\mathcal Ku_0u_1=q/(4r^2|X'|Q)$.

As a consequence, as it was already stated in \eqref{eq:exceptional-H-positivity}, for $q>0$, and on a regular interval where the background coefficients are nonvanishing, this Hamiltonian is non-negative for arbitrary compactly supported canonical data if and only if
\begin{equation}
\mathcal K>0\,,\qquad Q>0
 \qquad\Longleftrightarrow\qquad
 Q>0\,,\qquad (rX)'>0\,.
\end{equation}

\bibliographystyle{JHEPmod}
	\bibliography{refs}

\end{document}